\documentclass[acmsmall,screen,nonacm,dvipsnames]{acmart}
\usepackage{changepage}
\usepackage{graphicx}
\usepackage{natbib}
\usepackage{xspace}
\usepackage{listings}
\usepackage{hyperref}
\usepackage{soul}
\usepackage{float}
\usepackage{wrapfig}
\floatstyle{boxed} 
\restylefloat{figure}
\usepackage[scale=1.0]{FiraMono}
\usepackage[T1]{fontenc}

\newcommand{\ruleName}[1]{\textbf{\footnotesize #1}}
\newcommand{\metaDef}{\mathrel{\mathop:}=}
\newcommand{\cmid}{\textcolor{NavyBlue}{\ \mid\!\!\mid \ }}
\newcommand{\optional}[1]{\textcolor{NavyBlue}{[\!\![} #1 \textcolor{NavyBlue}{]\!\!]}}
\newcommand{\many}[5]{\textcolor{NavyBlue}{\left(\textcolor{black!100}{#1}\right)_{#3 \in \mathcal{\uppercase{#5}}}}}

\newcommand{\instRel}{\ \preceq \ }

\newcommand{\keyword}[1]{\textcolor{Violet}{\mathtt{#1}}}

\newcommand{\evalRel}{\rightarrowtail}
\newcommand{\stepRel}{\rightsquigarrow}

\newcommand{\cstOf}{\rhd}
\newcommand{\bunch}[1]{[{#1}]}
\newcommand{\scApp}{\Leftarrow}
\newcommand{\scAppCode}{\ }

\newcommand{\scFun}{\Rightarrow}
\newcommand{\sFun}[3]{{#2}\xrightarrow{#1}{#3}}

\newcommand{\tk}{\mathcal{T}}

\newcommand{\tdk}{\tk_\bullet}
\newcommand{\rk}{\mathcal{R}}

\newcommand{\nop}{\star}

\newcommand{\tSigma}{\overline{\sigma}}

\newcommand{\inc}{\downarrow}
\newcommand{\outc}{\uparrow}
\newcommand{\sRes}[2]{{#1} \blacktriangleright {#2}}
\newcommand{\alt}{\diamond}

\newcommand{\elevate}{ELEVATE\xspace}
\newcommand{\code}[1]{\texttt{\footnotesize #1}}

\lstdefinestyle{miniElevateStyle}{
    basicstyle=\ttfamily\footnotesize
}

\usepackage{amsmath}
\usepackage{cleveref}
\usepackage{amsfonts}
\usepackage{wasysym}
\usepackage{mathtools}

\usepackage{enumitem}
\setlistdepth{9}

\setlist[itemize,1]{label=$\bullet$}
\setlist[itemize,2]{label=$\star$}
\setlist[itemize,3]{label=$\ast$}
\setlist[itemize,4]{label=$\circ$}
\setlist[itemize,5]{label=$-$}
\setlist[itemize,6]{label=$\cdot$}
\setlist[itemize,7]{label=$\cdot$}
\setlist[itemize,8]{label=$\cdot$}
\setlist[itemize,9]{label=$\cdot$}

\renewlist{itemize}{itemize}{9}

\usepackage{tikz}

\usetikzlibrary{tikzmark}
\usetikzlibrary{calc}

\usepackage{scalerel}
\usepackage{comment}

\usepackage{booktabs}   
\usepackage{subcaption} 

\usepackage{proof-at-the-end}

\newenvironment*{lemmaE}[1][]{
  \begin{theoremEnd}[see full proof, end, restate]{lemma}[#1]
}{
  \end{theoremEnd}
}

\lstdefinelanguage{elevate}{
    morecomment=[l]{--}, 
    morecomment=[l]{//}, 
    morecomment=[s]{/*}{*/}, 
    alsoletter = {;, ->, ||, \#},
    keywords=[1]{
        let,
        in,
        lam,
        st,
        sc,
        run,
        rec,
        match,
        as,
        forall,
        return,
        seq,
        lChoice,
        rule,
        try,
        else, ;, ->, ||,
        choice,
        \#,
        succ,
        fail
    },
    keywords=[2]{
        Rise,
        Id,
        Name,
        Lam,
        Param,
        Body,
        App,
        Fun,
        Arg,
        Primitive,
        Map,
        Zip,
        Reduce,
        Op,
        Mult,
        Add,
        Lhs,
        Rhs,
        map,
        zip,
        reduce,
        fun,
        app,
    },
    keywords=[3]{
        RewriteResult,
        Success,
        success,
        Failure,
        Strategy,
        True,
        False,
        Boolean
    },
    keywords=[4]{
        f,
        g,
        h,
        x,
        y,
        z,
        xs,
        ys,
        m,
        n,
        u,
        v,
        w
    },
}
\setcopyright{none}

\acmJournal{PACMPL}
\acmVolume{1}
\acmNumber{OOPSLA} 
\acmArticle{1}
\acmYear{2023}
\acmMonth{1}
\acmDOI{} 
\startPage{1}

\begin{document}

\title{Traced Types for Safe Strategic Rewriting}

\author{Rongxiao Fu}
\affiliation{
  \institution{University of Edinburgh}           
  \country{United Kingdom}                   
}
\email{s1742701@ed.ac.uk}          

\author{Ornela Dardha}
\affiliation{
  \institution{University of Glasgow}           
  \country{United Kingdom}                   
}
\email{ornela.dardha@glasgow.ac.uk}         

\author{Michel Steuwer}
\orcid{0000-0001-5048-0741}             
\affiliation{
  \institution{The University of Edinburgh}           
  \country{United Kingdom}                   
}
\email{michel.steuwer@ed.ac.uk}         


\begin{abstract}
Strategy languages enable programmers to compose rewrite rules into \emph{strategies} and control their application.
This is useful in programming languages, e.g., for describing program transformations compositionally, but also in automated theorem proving, where related ideas have been studies with tactics languages.
Clearly, not all compositions of rewrites are correct, but how can we assist programmers in writing correct strategies?

In this paper, we present a static type system for strategy languages.
We combine a \emph{structural type system} capturing how rewrite strategies transform the shape of the rewritten syntax with a novel \emph{tracing system} that keeps track of all possible legal strategy execution paths.
Our type system raises warnings when parts of a composition are guaranteed to fail at runtime, and errors when no legal execution for a strategy is possible.

We present a formalization of our strategy language and novel tracing type system, and formally prove its type soundness.
We present formal results, showing that ill-traced strategies are guaranteed to fail at runtime and that well-traced strategy executions \emph{``can't go wrong''},
meaning that they are guaranteed to have a possible successful execution path.

\end{abstract}




\maketitle

\section{Introduction}
Rewrite systems find applications in many domains ranging from logic~\citep{DBLP:conf/alp/Marchiori94} and theorem provers~\citep{DBLP:journals/jlp/HsiangKLR92} to specifying the semantics of programming languages~\cite{DBLP:journals/jlp/RosuS10} and program transformation in compilers~\citep{VISSER2001109}.
In many domains, it is sufficient to simply specify a set of \emph{rewrite rules}--each specifying an individual rewrite step--which are applied (possibly non-deterministically) by a rewriting system until a normal form is reached or no rule is applicable anymore.

But in many domains this is not adequate.
For example, \emph{rewrite rules} are a straightforward choice for encoding simple program transformation, but are not sufficient to encode more complex program transformations and for controlling their applications in practice.
It might be required to apply a rule only to a subpart of a program, apply multiple rules in a specific order, or en-/disable rules during a specific phase.
To control the application of rewrite rules, \emph{strategy languages}, such as Stratego \cite{visser1998building,martin2008stratego} have been proposed.
\citet{kirchner2015rewriting} provides a recent overview of the field.
These strategy languages enable \emph{strategic rewriting} by providing combinators to compose rewrite rules into larger strategies.

Stratego is an integral part of the Spoofax language workbench by \citet{DBLP:conf/oopsla/KatsV10a} designed to declaratively specify languages and tailored IDEs to work with them.
The Stratego strategy language is used here to write interpreters and compilers purely using compositions of rewrites.

\citet{bastianhagedorn2020achieving,DBLP:journals/cacm/HagedornLKQGS23} describes the ELEVATE strategy language and how it is used to encode and control the application of traditional compiler optimizations, such as loop-tiling, compositionally.
The achieved performance is comparable to the traditionally designed TVM compiler~\citep{chen2018tvm} for deep learning.
In fact, ELEVATE shows how to rethink the design of \emph{``user-schedulable languages''} as strategy languages, as highlighted by \citet{DBLP:journals/cacm/RaganKelley23}.
\emph{Schedules} allow experts precise control over what compiler optimizations to apply, an idea popularized by the domain-specific compiler Halide \citep{DBLP:conf/pldi/Ragan-KelleyBAPDA13} and now widely adopted in other optimizing compilers, such as TVM.

Closely related to strategy languages are \emph{tactic languages} in automatic theorem proving that allow to control the arrangement of individual proof steps.
These ideas go all the way back to \citet{DBLP:conf/popl/GordonMMNW78} and tactic languages are now found in many proof assistants, such as in Coq \cite{DBLP:conf/lpar/Delahaye00}.

Strategy languages, such as Stratego and ELEVATE, empower programmers to describe complex strategies as compositions of rewrites.
But, clearly, not all compositions of rewrites are correct.
As a simple example, consider the sequential composition (\lstinline!;!) of two rewrite \lstinline[basicstyle=\normalsize]!rule!s:
\begin{lstlisting}[frame=none]
  let broken_composition = rule x + y -> y + x ; rule 1 * z -> z
\end{lstlisting}

In this composition, we first apply a commutativity rewrite, before attempting to remove the 1 as the neutral element of multiplication.
But clearly, something has gone wrong here!
The first rewrite will produce an expression with an addition operator, but the second rewrite expects a multiplication operator.
This composition of rewrites can never work and will fail at runtime for any possible input.
While this is a simple example, it raises a general question:
how do we assist programmers to avoid such mistakes and in writing correct composition of rewrites as strategies?
\medskip

In this paper, we present a novel powerful static type system for strategy languages, that is capable of rejecting ill-composed strategies such as the example from above.
To achieve this, we encode the syntax of rewritten expressions (such as \lstinline!x+y!) as structural types using row-polymorphic variant types.
We then give rewrite rules and their compositions as strategies function-like \emph{strategy types} reflecting the syntactic program transformation described by the rewrite.
When applying a strategy to an input, our type system gives the resulting value a \emph{strategy result type} that reflects that the rewrite is either performed successfully, resulting in the rewritten expression, or failed to apply, resulting in a runtime \emph{failure}.
Finally, to account for higher-order combinators that take strategies as arguments we give them dedicated \emph{strategy combinator types}.
We combine the structural type system with a novel \emph{tracing type system} to statically check that a successful execution path exist for a composition of rewrites, rejecting compositions of rewrites that are guaranteed to results in a runtime failure for all possible inputs.
Our tracing type system guarantees this not just for sequential compositions, as seen above, but also in the presence of the choice combinator (\lstinline!||!) that picks among two possible rewrite strategies.
\medskip

We present a formalization of our statically typed core strategy language \emph{Typed \elevate} with combinators similar to the once found in ELEVATE (and Stratego).
We present the language's novel type system that combines structural types with traces and prove its type soundness.
Furthermore, we present formal results showing that strategies which are structurally well-typed and well-traced are free of composition errors, as the one in the example above, and, with suitable input their execution is guaranteed to have a possible successful execution path.
On the contrary, we formally show that structurally well-typed strategies which are \emph{not} well-traced \emph{must} result in a runtime failure and, therefore, are justifiably rejected by our type system.
\medskip

To summarize, this paper makes the following contributions:
\begin{itemize}
  \item we introduce a powerful static type systems for strategy languages combining structural types and the novel traced types for rejecting ill-composed strategies;\\[-.5em]
  \item we discuss a number of examples demonstrating the type system in practice  (\Cref{sec:by-example});\\[-.5em]
  \item we present a formalization of the statically typed core strategy language \emph{Typed \elevate} and its type system (\Cref{sec:term-syntax}--\ref{sec:aux-typing}) and prove its type soundness (\Cref{sec:op-semantics}--\ref{sec:type-prop});\\[-.5em]
  \item we formally define what it means for well-typed and well-traced strategies \emph{``not to go wrong''} and we show that well-typed but \emph{not} well-traced strategies \emph{must} result in a runtime failure (\Cref{sec:st-prop}).
\end{itemize}
We discuss limitations and design choices in \cref{sec:discussion}, before discussing related work in \cref{sec:related-work} and concluding in \cref{sec:conclusion}.

\section{Typed \elevate by Example}
\label{sec:by-example}
\label{sec:prog}

In this section, we present Typed \elevate by example, starting in \cref{sec:valid_compositions} with discussing valid compositions of rewrites showing the types given to these strategies by our type system.
We will then discuss problematic compositions which our type system will flag with warnings in \cref{sec:compositions_with_warnings} and compositions that our type system rejects with errors in \cref{sec:compositions_with_errors}.

\subsection{Valid strategy compositions, strategy applications, and strategy combinators}
\label{sec:valid_compositions}

\subsubsection{Example 1}

Let's start with a simple composition of three rewrite rules:
\begin{lstlisting}[mathescape]
let e1 = rule m * n -> n * m ; (rule 1 * v -> v || rule 2 * w -> w + w)
\end{lstlisting}
To define the strategy \lstinline!e1!, we first swap the two factors \lstinline!n! and \lstinline!m!, before either removing an unnecessary multiplication with \lstinline!1! or, alternatively, turning a multiplication with \lstinline!2! into an addition.
We use the sequential composition operator (\lstinline!;!) that applies the second strategy to the output of the first, and the choice combinator (\lstinline!||!) that applies one of two strategies, or fails if none is applicable.

To represent the expressions we rewrite, such as \lstinline!m * n!, as types we encode them using row-polymorphic variant types, as formally explained in \cref{sec:formal}.
In this section, we will write these structural types informally as we would present them to users of the type system.
Rewrite rules and their compositions as strategies are given a \emph{strategy type} that reflects the syntactic transformation described by the rewrite.
For example, the first rewrite rule has the following strategy type:
\begin{lstlisting}[mathescape, frame=none, aboveskip=0.125em, belowskip=0.25em]
  rule m * n -> n * m : $\sFun{\bunch{a}}{a_0 * a_1}{a_1 * a_0}$
\end{lstlisting}
The variables appearing in the type are \emph{trace variables}.
Here all variables belong to a single trace identified by $a$ indicating that there is only a single possible way to perform the rewrite described by the strategy:
swapping the two operands of the multiplication, represented by the two trace variables $a_0$ and $a_1$.


The strategy type for the choice composition of the two other rewrites is more interesting:
\begin{lstlisting}[mathescape, frame=none, aboveskip=0.125em, belowskip=0.25em]
  (rule 1 * v -> v || rule 2 * w -> w + w) : $\sFun{\bunch{b,c}}{ 1 * b_0 ~|~ 2 * c_0 }{ b_0 ~|~ c_0 + c_0}$
\end{lstlisting}
Here the strategy type has two traces, $b$ and $c$, indicating that there are two possible executions of this rewrite strategy:
either, the first rewrite rule is performed transforming $1*b_0$ into $b_0$,
or, the second rewrite rule transforms the multiplication $2*c_0$ into $c_0 + c_0$.
When there are multiple traces in a strategy type, we write the input and output for each trace individually separated by $|$ to indicate that these are separated possible executions of the strategy.
In our formal system, types are not represented this way.
We present a justification in \cref{sec:formal} why this simplified and intuitive presentation of types is a truthful reflection of the information available in the formal type.

When sequentially combining both strategies, we obtain a strategy type that reflects the effect of the overall strategy:
\begin{lstlisting}[mathescape]
let e1 = rule m * n -> n * m ; 
         (rule 1 * v -> v || rule 2 * w -> w + w) : $\sFun{\bunch{d,e}}{ d_0 * 1 ~|~ e_0 * 2 }{ d_0 ~|~ e_0 + e_0 }$
\end{lstlisting}
The strategy type now indicates that there are two possible ways to successfully perform the composed strategy:
either the input expression has the form $d_0*1$ and it will be rewritten into $d_0$,
or, alternatively, the input has the form $e_0*2$ and it will be rewritten into $e_0+e_0$.
For any other expression as input the composed strategy will fail.

\subsubsection{Example 2}
Can we also check the application of rewrite strategies to input expressions?
Consider the following example where we pass the expression \lstinline!5 * 2! as input to the strategy \lstinline!e1! from \emph{Example 1} and then pass the resulting expression as input to an additional rewrite rule:

\begin{lstlisting}[mathescape]
let e2 = (rule 5 + 5 -> 10) (e1 (5 * 2))
\end{lstlisting}

The application of \lstinline!5 * 2! to \lstinline!e1! has the \emph{strategy result type}: $\sRes{\bunch{b}}{5 + 5}$.
In this case it indicates a successfully rewritten expression, written on the right side of the black triangle.
No trace variable $b_i$ appears on the right side, as the expression only contains constants and no variables.

With this type as the input for the rule \lstinline[mathescape]!(rule 5 + 5 -> 10)  : $\sFun{\bunch{a}}{5 + 5}{10}$!, we obtain:
%
%
\begin{lstlisting}[mathescape]
let e2 = (rule 5 + 5 -> 10) (e1 (5 * 2)) : $\sRes{\bunch{c}}{10}$
\end{lstlisting}

This type confirms that this strategy execution is valid and will produce the expression \lstinline!10!.

\subsubsection{Example 3}
Next, let's consider a sequential composition of two choice compositions:
\begin{lstlisting}[mathescape]
let e3 = (rule m * n -> n * m || rule m + n -> n + m) ;
         (rule 1 * v -> v     || rule 0 + w -> w)
\end{lstlisting}
Here, the choice composition in the first line swaps the operands of multiplication or addition, the second choice composition in the second line removes the identity element for multiplication or addition.
When sequentially composing two choice compositions in general there are four possible execution paths: left-left, left-right, right-left, and right-right.
But clearly, in this example only two will execute successfully:
the first one swapping the operands and removing the neutral element for multiplication (left-left) and the other for addition (right-right).
To reflect this, this example could also have been written as:
\begin{lstlisting}[mathescape, frame=none, aboveskip=0.125em, belowskip=0.25em]
  (rule m * n -> n * m ; rule 1 * v -> v) || (rule m + n -> n + m ; rule 0 + w -> w)
\end{lstlisting}
%
%
Our tracing type system is capable of precisely reflecting all possible executions paths in the type.
The individual choice compositions have the types:
\begin{lstlisting}[mathescape, frame=none, aboveskip=0.125em, belowskip=0.25em]
  (rule m * n -> n * m || rule m + n -> n + m) : $\sFun{\bunch{a, b}}{a_0*a_1 ~|~ b_0+b_1}{a_1*a_0 ~|~ b_1+b_0}$
\end{lstlisting}
and
\begin{lstlisting}[mathescape, frame=none, aboveskip=0.125em, belowskip=0.25em]
  (rule 1 * v -> v     || rule 0 + w -> w)     : $\sFun{\bunch{c, d}}{1*c_0 ~|~ 0+d_0}{c_0~|~d_0}$
\end{lstlisting}

When sequentially composing them, our tracing type system connects the traces of both strategy types to compute the possible traces of the composition representing all possible execution paths.
\begin{lstlisting}[mathescape]
let e3 = (rule m * n -> n * m || rule m + n -> n + m) ;
         (rule 1 * v -> v     || rule 0 + w -> w) : $\sFun{\bunch{e,f}}{e_0 * 1 ~|~ f_0 + 0}{e_0~|~f_0}$
\end{lstlisting}
The final type of the composition clearly expresses that the overall strategy rewrites either $e_0*1$ into $e_0$ or $f_0+0$ into $f_0$.
The strategy must fail for all other inputs.

We will discuss in \Cref{sec:formal} formally how our type system computes the connected traces of strategy compositions.

\subsubsection{Example 4}
So far we have only seen strategies that are build using the two fundamental combinators \lstinline!;! and \lstinline!||!.
But, we can build custom strategy combinators as well, these take strategies as arguments and augment their behavior.
For example, let's consider the \lstinline!swapOps! combinator that takes a strategy \lstinline!s! and applies it after swapping the operands of an arbitrary binary operator \lstinline!op!.
\begin{lstlisting}[mathescape]
let swapOps = st s => rule m op n -> n op m ; s
\end{lstlisting}

This custom combinator can be used to swap the order of operands for any strategy transforming binary expressions, such as: \lstinline[mathescape]!swapOps (rule 1 * v -> v)! which has the type: \lstinline[mathescape]!$\sFun{\bunch{b}}{b_0 * 1}{b_0}$!.

Strategy combinators have a dedicated arrow type $\scFun$.
For our custom combinator this is:
\begin{lstlisting}[mathescape]
let swapOps = st s => rule m op n -> n op m ; s : $(\sFun{\bunch{a}}{(a_0 \ a_2 \ a_1)}{a_3}) \scFun (\sFun{\bunch{a}}{(a_1 \ a_2 \ a_0)}{a_3})$
\end{lstlisting}

This type indicates that the strategy input expression changes its order from $a_0\ a_2\ a_1$ to $a_1\ a_2\ a_0$, clearly showing the swap of the operands.

\subsection{Problematic Strategy Compositions resulting in Warnings}
\label{sec:compositions_with_warnings}
After a number of examples showing valid strategies, we will now investigate problematic strategies.
In this section, we investigate an example that will still execute correctly, but contains a dead execution branch that our type system detects and warns the user about.

\subsubsection{Example 5}
Let's consider the following strategy:
\begin{lstlisting}[mathescape]
let e5 = (rule m + n -> n + m ; rule m * n -> n * m) || rule 1 * v -> v
\end{lstlisting}

This is a choice composition of a sequential composition on the left and a rewrite rule on the right.
The sequential composition is clearly problematic, as the first rewrite rule produces an expression incompatible with the second rule.
Our type system will give the sequential composition a strategy type with an \emph{empty trace} (\lstinline[mathescape]!$\sFun{\bunch{}}{\_}{\_}$!) which indicates that there is no possible execution path.

But, the overall strategy \lstinline!e5! will still work fine when using the rule on the right.
Therefore, in this case the overall composition has the same type as the rule on the right:
\lstinline[mathescape]!$\sFun{\bunch{a}}{1 * a_0}{a_0}$!.
This shows that the composed strategy will work for expressions that are inputs for the rule on the right.

To alert the user of the problematic sequential composition, an implementation of our type system could issue a warning to the user:
\begin{lstlisting}[mathescape]
let e5 = (rule m * n -> n * m ; rule m + n -> n + m) || rule 1 * v -> v : $\sFun{\bunch{a}}{1 * a_0}{a_0}$
//       ^~~~~~~~~~~~~~~~~~~~~~~~~~~~~~~~~~~~~~~~~~~
//       Warning: this strategy is guaranteed to fail at runtime,
//                but the overall strategy can still succeed
\end{lstlisting}

In \Cref{sec:formal} we will see the typing rule $\ruleName{T-S-App}$ that could raise these warnings as soon as the set of traces in the conclusion becomes empty.

\subsection{Problematic Strategy Compositions resulting in Errors}
\label{sec:compositions_with_errors}
Finally, in this section, we discuss examples of problematic strategies that are guaranteed to fail at runtime for all possible inputs and are, therefore, rejected by our type system.

\subsubsection{Example 6}
Let's consider a problematic sequential composition first:
\begin{lstlisting}[mathescape]
let e6 = rule 2 * n -> n + n ; rule 2 + 3 -> 5
\end{lstlisting}

On a quick glance this might seem like a valid composition, as the first rule produces a multiplication expression which is expected by the second rule.
But, clearly we can not instantiate the same variable \lstinline!n! with \lstinline!2! and \lstinline|3| at the same time!
Formally in our type system this will be detected as there will be no trace for the unified typed.
This is a similar situation for the example before where we reported only a warning.
How do we distinguish when raising a warning and when rejecting a program with an error?
It depends on the context in which an empty traced strategy is used.
If it is used in a larger composition (as in the example before) we will only raise a warning, but if we try to bind an empty traced strategy to a name using \lstinline!let! we raise an error:
\begin{lstlisting}[mathescape]
let e6 = rule 2 * n -> n + n : $\sFun{\bunch{a}}{2 * a_0}{a_0 + a_0}$
       ; rule 2 + 3 -> 5     : $\sFun{\bunch{b}}{2 + 3}{5}$
//     ^~~~~~~~~~~~~~~~~~~~~~~
//     Error: this sequential composition is guaranteed to fail at runtime
//            resulting in a guaranteed failure of the overall strategy
//            
//            There is no trace for the composed strategy type
//            after unification of $a_0 + a_0$ and $2 + 3$
\end{lstlisting}

%

\subsubsection{Example 7}
Finally, let us consider a more slightly complicated example with a sequential composition of two choice compositions:
\begin{lstlisting}[mathescape]
let e7 = (rule m * n -> n * m || rule m + n -> n + m) ;
         (rule v - 0 -> v     || rule w / 1 -> w)
\end{lstlisting}

Here, no successful execution is possible as the possible expressions produced by the first strategy must contain a multiplication or addition which are incompatible with the second strategy expecting subtraction or division.
We can give each choice composition a valid strategy type, but we can't find a trace for the composition after unifying the output type of the first strategy with the input type of the second:
\begin{lstlisting}[mathescape]
let e7 = (rule m * n -> n * m) || (rule m + n -> n + m) : $\sFun{\bunch{a,b}}{a_0*a_1 ~|~ b_0+b_1}{a_1*a_0 ~|~ b_1+b_0}$
       ; (rule v - 0 -> v      || rule w / 1 -> w)      : $\sFun{\bunch{c,d}}{c_0-0 ~|~ d_0 / 1}{c_0~|~d_0}$
//     ^~~~~~~~~~~~~~~~~~~~~~~~~~~~~~~~~~~~~~~~~~~~~~~~~~
//     Error: this sequential composition is guaranteed to fail at runtime
//            resulting in a guaranteed failure of the overall strategy
//            
//            There is no trace for the composed strategy type
//            after unification of $a_1*a_0 ~|~ b_1+b_0$ and $c_0-0 ~|~ d_0/1$
\end{lstlisting}

In \Cref{sec:formal} we will see the typing rule $\ruleName{T-Let}$ that will fail with an error for empty traced strategies.
\medskip

In this section, we have seen a number of examples showing the types we assign to strategies and strategy combinators.
We have seen successfully composed strategies as well as problematic ones that are highlighted with a warning or rejected with an error.
Crucially, \emph{traces} indicate the possible execution paths that can be taken at runtime by a strategy.
Problematic strategies are identified, when the set of traces is empty indicating that no possible execution path exists.
In the next section, we formalize these intuitions and formally show that well-typed but empty-traced strategies \emph{must} fail at runtime, justifying why our type system rejects such strategies.
Furthermore, we will prove that our type system is sound, and formally clarify what it means for well-typed and well-traced strategies to \emph{``not go wrong''}.

\section{Typed \elevate, Formalized}
\label{sec:formal}
In this section we present the formalization of Typed \elevate.
Unless otherwise stated, elements in \(\textcolor{NavyBlue}{NavyBlue}\) are meta-level descriptions, for e.g., the square brackets \(\optional{\ }\) indicate \emph{option} in the EBNF grammar; the indexed multiple occurrences (possibly separated by \(\ ,\ \)) of a syntactical construct \(expr\) are collectively represented by \(\many{expr_i}{,}{i}{0}{n}\), where the index (written as \(i\), \(j\), \(k\), \(p\) or \(q\)) ranges over a possibly empty subset (written as \(\mathcal{M}\), \(\mathcal{N}\), \(\mathcal{U}\) or \(\mathcal{V}\)) of the set of natural numbers. When two index sets occur in the same formula, they are disjoint by default.

\subsection{Term Syntax}
\label{sec:term-syntax}

\begin{figure}
  \footnotesize
    \begin{alignat*}{4}
      \mathrm{Patterns} \quad
      &p \ &\metaDef \quad & x,y,z,\dots \cmid () \cmid (p_m, p_n) \cmid \ell \ p \\
      \mathrm{Terms} \quad
      &e \ &\metaDef \quad & x,y,z,\dots \cmid \keyword{rule} \ p \to e \cmid () \cmid (e_m, e_n) \cmid \ell \ e \cmid\\
      &&& \keyword{seq} \cmid \keyword{choice} \cmid \keyword{st} \ x \scFun e_b \cmid e_f \scApp e \cmid \\
      &&&e_s \gets e_i \cmid \keyword{succ} \ e \cmid \keyword{fail} \cmid e_m \alt e_n \cmid \\
      &&&\keyword{let} \ x = e_f \ \keyword{in} \ e
    \end{alignat*}
  \caption{Syntax of terms and patterns.}
  \label{fig:ast}
  \label{fig:syntax}
\end{figure}
\Cref{fig:syntax} shows the syntax of terms and patterns.

\paragraph{Patterns}
Patterns (denoted by $p$) are used in rewrite rule definitions, include pattern variables ranged over valid variable names (denoted by $x,y,z, \cdots$), unit pattern ($()$), pair pattern ($(p, p)$), and variant pattern ($\ell \ p$) where $\ell$ is the label to be matched. For simplicity, all patterns in this paper are \emph{linear}, namely each bound pattern variable only appears once in a pattern. The linearity requirement is enforced by typing rules for patterns in \Cref{fig:rule-typing}.

\paragraph{Terms}
Terms (denoted by $e$) include variables ranged over valid variable names (denoted by $x,y,z, \cdots$). As the basic component of the language, rewrite rules can be defined with the lambda-like syntax ($\keyword{rule} \ p \to e$), where $p$ is the pattern to be matched or the LHS (Left-hand Side) of the rule, and $e$ is the RHS (Right-hand Side) of the rule constructed with the variables from $p$ and other syntax components. Thus, the term syntax provides unit constructor ($()$), pair constructor ($(e_m, e_n)$), and variant constructor or label application ($\ell \ e$) for use inside rule definitions. It is straightforward to see that the syntax of these terms is similar to that of patterns and a conversion $\mathbf{p2e}$ from patterns to terms can be easily defined (see \Cref{fig:p2e} in the appendix). There are two constant strategy combinators, $\keyword{seq}$ for sequential composition and $\keyword{choice}$ for the (non-deterministic) choice composition of strategies. To allow the programmers to define strategy combinators by themselves, the term syntax also provides strategy combinator abstraction ($\keyword{st} \ x \scFun e_b$) where $x$ is the bound variable representing a strategy and $e_b$ is the body of the combinator definition, and strategy combinator application ($e_f \scApp e$) where $e_f$ is the strategy combinator and $e$ is the argument strategy. To get the execution result of strategies, one can apply an input $e_i$ to a strategy $e_s$ by writing strategy execution ($e_s \gets e_i$), and the execution result can either be ($\keyword{succ} \ e$) where $e$ is the successfully rewritten expression, or simply $\keyword{fail}$. To support the $\keyword{choice}$ combinator, non-deterministic choice between two results ($e_m \alt e_n$) is also involved in the syntax, but in most cases, programmers do not need to use this operator directly. Finally, the term syntax provides let-bindings ($\keyword{let} \ x = e_f \ \keyword{in} \ e$) for polymorphic abstractions of strategies, strategy combinators and execution results.

\paragraph{Informal Presentation of Terms}
Shown below is a term written in the formal syntax. It contains a strategy combinator definition (\lstinline{swapOps} from Example 4) and strategy applications.
\begin{lstlisting}[mathescape,frame=none]
$\keyword{let} \ swapOps = \keyword{st} \ s \scFun \keyword{seq} \scApp (\keyword{rule} \ Op \ (op, (m, n)) \to Op \ (op, (n, m))) \scApp s$
$\keyword{in} \ (swapOps \scApp (\keyword{rule} \ Op \ (Mul \ (), (1 \ (), v)) \to v)) \gets \keyword{succ} \ (Op \ (Mul \ (), (2 \ (), 1 \ ())))$
\end{lstlisting}

By applying the syntactic sugar for Typed \elevate terms such as numbers and binary operators in rewrite rule definitions, we get a term in the simplified syntax used in \Cref{sec:by-example}.
\begin{lstlisting}[mathescape,frame=none]
let swapOps = st s => (rule m op n -> n op m) ; s in swapOps (rule 1 * v -> v) (2 * 1)
\end{lstlisting}

\subsection{Type Syntax}
\label{sec:type-syntax}

\begin{figure}
  \footnotesize
  \centering
  \begin{minipage}{.45\textwidth}
    \begin{alignat*}{4}
      \mathrm{Trace \ Variables} \quad 
      &\gamma \ &\metaDef \quad &\alpha \cmid \beta \\
      \mathrm{Trace \ Identifier \ Sets} \quad 
      &\varphi \ &\metaDef \quad &\bunch{\many{\alpha_i}{,}{i}{0}{n}} \\
      \mathrm{Trace \ Member \ Sets} \quad 
      &\phi \ &\metaDef \quad &\bunch{\many{\beta_i}{,}{i}{0}{n}} \\
      \mathrm{Trace \ Variable \ Sets} \quad 
      &\psi \ &\metaDef \quad &\bunch{\many{\gamma_i}{,}{i}{0}{n}} \\
      \mathrm{Traces} \quad 
      &&\alpha \cstOf \phi
    \end{alignat*}
    \begin{alignat*}{4}
      \mathrm{Kinds} \quad 
       &\kappa \ &\metaDef \quad &\rk \cmid \tk \cmid \tdk\\
       &\mathcal{R} \ &\metaDef \quad &\{\many{l_i}{,}{i}{0}{n}\}
    \end{alignat*}
    \begin{alignat*}{4}
      \mathrm{Types} \quad 
      &t \ &\metaDef \quad &\tau \cmid \rho \cmid \omega\\
      &\tau \ &\metaDef \quad &\nu_{\optional{\phi}} \cmid \langle \rho \rangle_{\optional{\phi}} \cmid (\nu \ \keyword{as} \ \langle \rho \rangle)_{\optional{\phi}} \cmid ()_{\optional{\psi}} \cmid (\tau_m, \tau_n)_{\optional{\phi}}\\
      &\rho \ &\metaDef \quad &\nu \cmid \cdot \cmid \ell: \tau \mid \rho\\[-.25em]
      &\omega \ &\metaDef \quad &\sRes{\varphi}{\tau} \cmid \sFun{\varphi}{\tau_p}{\tau_e} \cmid (\sFun{\varphi}{\tau_p}{\tau_e}) \scFun \omega
    \end{alignat*}
  \end{minipage}%
  \hfill\vline\hfill%
  \begin{minipage}{.4\textwidth}
    \begin{alignat*}{4}
      \mathrm{Schemes} \quad 
      &\tSigma \ &\metaDef \quad &\keyword{tr} \ \Phi. \ \sigma
      \\
      &\sigma \ &\metaDef \quad &\forall \ (\nu: \kappa). \ \sigma \cmid \omega
      \end{alignat*}
    \vspace{1em}\hrule\vspace{1em}
    \begin{alignat*}{4}
      \mathrm{Kinding \ Env.} \quad 
      &\Delta \ &\metaDef &\cdot \cmid \Delta, \nu: \kappa
      \\
      \mathrm{Typing \ Env.} \quad 
      &\Gamma \ &\metaDef &\cdot \cmid \Gamma, x: \tSigma
      \\
      \mathrm{Tracing \ Env.} \quad 
      &\Phi \ &\metaDef \ &\cdot \cmid \Phi, \alpha \cstOf \phi
      \\
      \mathrm{Rule \ Env.} \quad 
      &\Theta \ &\metaDef &\cdot \cmid \star \cmid \Theta, x: \tau
      \\
      \mathrm{Strategy \ Combinator \ Env.} \quad 
        &\Sigma \ &\metaDef \ &\cdot \cmid \Sigma, x: \omega
      \end{alignat*}
  \end{minipage}

\caption{Syntax of Traces, kinds and types (left) and kinding, tracing and typing environments (right).}
\label{fig:traces}
\label{fig:types}
\end{figure}

\Cref{fig:types} shows the syntax of traces, kinds, and types in Typed \elevate, together with the syntax of the kinding, tracing and typing environments.

\paragraph{Traces}
A novel and important part of our type system is \emph{tracing}. With traces, we are able to observe each possible way of rewriting at the type level. Traces are constructed with trace variables ($\gamma$) which can be partitioned into two subsets: trace identifiers ($\alpha$) and trace members ($\beta$). We use different naming conventions for different kinds of trace variables in the examples, and different meta variables to represent them in our formalism for convenience and clarity. Notice however that it is not necessary for trace variables to have any syntactical difference because the tracing environment can distinguish them. The sets (of distinct elements) formed by different group of trace variables are also denoted by different meta variables: $\varphi$ for trace identifier sets, $\phi$ for trace member sets, and $\psi$ for trace variable sets. Finally, we are able to define traces. Each trace denoted as $\alpha \cstOf \phi$ is identified by the trace identifier $\alpha$ and the trace member set $\phi$ contains the remaining trace variables of this trace. Besides adapting common set operations, for convenience, the following conversion rules are used for writing the sets mentioned above. We show the rules for trace identifier sets as an example, and the same rules can be applied to the other two kinds of sets.
\begin{center}
\code{$\bunch{\alpha, \varphi} = \bunch{\alpha, \many{\alpha_i}{,}{i}{0}{n}} \ if \ \varphi = \bunch{\many{\alpha_i}{,}{i}{0}{n}} \ and \ \alpha \notin \varphi$}

\code{$\bunch{\varphi_a, \varphi_b} = \bunch{\many{\alpha_i}{,}{i}{0}{n}, \many{\alpha_j}{,}{j}{0}{m}} \ if \ \varphi_a = \bunch{\many{\alpha_i}{,}{i}{0}{n}} \ and \ \varphi_b = \bunch{\many{\alpha_j}{,}{j}{0}{m}} \ and \ \varphi_a \cap \varphi_b = \emptyset$}
\end{center}

\paragraph{Kinds and Types}
There are three syntactical categories of types together forming the type universe $t$ in Typed \elevate: traceable types ($\tau$), row types ($\rho$), and traced types ($\omega$). Traceable types ($\tau$) are of kind $\tdk$, and they include type variables ($\nu$), variant types ($\langle \rho \rangle$) constructed from rows, equi-recursive variant types ($\nu \ \keyword{as} \ \langle \rho \rangle$) where $\nu$ is a variable representing the type itself, unit type ($()$), and pair types ($(\tau_m, \tau_n)$). For each case in the definition of $\tau$, there is a trace variable set (or trace member set) optionally attached to it at the subscript position. This is how tracing works at the syntactical level: instead of being deeply tied to the underlying types, traces are another layer of type-level information overlaid on the underlying types.

The following conversion rule is used for folding and unfolding equi-recursive variant types, where $\mathbf{erase}$ is a function erasing all traces and later defined in \Cref{fig:trace-erasure} in the appendix.
\begin{center}
\code{$\nu \ \keyword{as} \ \langle \rho \rangle \simeq \langle \rho \rangle[\nu \mapsto \mathbf{erase}(\nu \ \keyword{as} \ \langle \rho \rangle)]$}
\end{center}
Since all traces in the folded type are erased during unfolding, the traces stay the same before and after this conversion. If the case for the equi-recursive variant types is absent from subsequent definitions, it means that all equi-recursive variant types are unfolded in advance.

Row types (\(\rho\)) are sequences of label-type pairs $\ell: \tau$ ending with row variables ($\nu$) or empty row (\(\cdot\)), where the order of the label-term pairs is insignificant, and labels are all distinct. The set of all possible labels and the set of all possible variable names are disjoint. Rows are differentiated from ordinary types by their kinds, named row kinds (\(\mathcal{R}\)), which are sets of labels that are not present in the rows. For a row variable, its kinds describes the finite set of labels that the instantiation of the row variable \emph{must not} contain \cite{blume06cases, daniel2016liberating}. Although row types are used to constructed traceable types, row types themselves are not traceable. This indicates that we can switch between different representations of rows, or even different specifications of structural types in general, and add the tracing mechanism to it.

Traced types ($\omega$) contains strategy result types ($\sRes{\varphi}{\tau}$), strategy types ($\sFun{\varphi}{\tau_p}{\tau_e}$), and strategy combinator types ($(\sFun{\varphi}{\tau_p}{\tau_e}) \scFun \omega$). Traced types and traceable types together form the $\tk$ kind, which is the super kind of $\tdk$. As its name suggests, each case of the traced types carries one or more trace identifier sets $\varphi$, identifying the traces in the corresponding traceable types. The strategy result type $\sRes{\varphi}{\tau}$ is the type of the result of strategy executions which could be either $\keyword{succ} \ e$ or $\keyword{fail}$. In the $\keyword{succ} \ e$ case, $\tau$ in $\sRes{\varphi}{\tau}$ gives the type of $e$. Thus, the strategy result type can be considered as an analog of the \code{Maybe a} type in Haskell. The strategy type $\sFun{\varphi}{\tau_p}{\tau_e}$ is the type of strategies which are either individual rewrite rules or large strategies composed from smaller strategies. For a single rewrite rules, the type can be read as "a rule rewriting a term of type $\tau_p$ to a term of type $\tau_e$", and for more general strategies, the traces in $\tau_p$ and $\tau_e$ can be more complex and each trace marks one possible way of rewriting. The strategy combinator types $(\sFun{\varphi}{\tau_p}{\tau_e}) \scFun \omega$ is the type of strategy combinators which are either provided by the language as primitives such as $\keyword{seq}$ and $\keyword{choice}$ or defined by programmers using the $\keyword{st}$ keyword. A subtlety here is that instead of directly taking a strategy of type $\sFun{\varphi}{\tau_p}{\tau_e}$ as its argument, a strategy combinator of type $(\sFun{\varphi}{\tau_p}{\tau_e}) \scFun \omega$ expects a strategy whose type is \emph{compatible} with $\sFun{\varphi}{\tau_p}{\tau_e}$ and return a traced type by applying the result of trace computations to $\omega$. More about the trace computation will be introduced in \Cref{sec:st-typing}. Finally, the trace identifier sets in traced types are written out only for clarity, and they can actually be computed using the tracing rules which will be introduced in \Cref{sec:well-formed}.

\paragraph{Informal Presentation of Types}
The expression \lstinline[breaklines=true]{(rule 1 * v -> v || rule 2 * w -> w + w)} (from Example 1) is used here as an example to discuss how the formal types relate to the informal presentation used in \Cref{sec:by-example}.
The formal term syntax for the expression is:

\begin{lstlisting}[mathescape,frame=none]
$\keyword{choice} \scApp (\keyword{rule} \ Op \ (Mul \ (), (1 \ (), v)) \to v) \scApp (\keyword{rule} \ Op \ (Mul \ (), (2 \ (), w)) \to Op \ (Add \ (), (w, w)))
$
\end{lstlisting}

Its formal type is as follows:
\begin{lstlisting}[mathescape]
$\langle Op: (\langle Mul: ()_{\bunch{b, c}} \mid \nu_a \rangle, (\langle 1: ()_{\bunch{b}} \mid 2: ()_{\bunch{c}} \mid \nu_c \rangle, (\nu \ \keyword{as} \ \langle Op: (\langle Add: () \mid \nu_e \rangle,(\nu, \nu)) \mid \nu_d \rangle)_{\bunch{b_0, c_0}})) \mid \nu_b \rangle \sFun{\bunch{b,c}}{}{}$
$\langle Op: (\langle Add: ()_{\bunch{c}} \mid \nu_e \rangle,$
$\quad ((\nu \ \keyword{as} \ \langle Op: (\langle Add: () \mid \nu_e \rangle,(\nu, \nu)) \mid \nu_d \rangle)_{\bunch{c_0}}, (\nu \ \keyword{as} \ \langle Op: (\langle Add: () \mid \nu_e \rangle,(\nu, \nu)) \mid \nu_d \rangle)_{\bunch{c_0}})) \mid \nu_d \rangle_{\bunch{b_0}}$
\end{lstlisting}

There are two trace identifiers, $b$ and $c$, corresponding to the two rules composed by the $\keyword{choice}$ combinator. We focus on the types traced by trace variables identified by $b$, which are $b$ and $b_0$, and perform a bottom-up traversal of the whole type while ignoring the $c$ trace, we get the type:
\begin{lstlisting}[mathescape]
$\langle Op: (\langle Mul: ()_{\bunch{b}} \mid \nu_a \rangle, (\langle 1: ()_{\bunch{b}} \mid 2: () \mid \nu_c \rangle, (\nu \ \keyword{as} \ \langle Op: (\langle Add: () \mid \nu_e \rangle,(\nu, \nu)) \mid \nu_d \rangle)_{\bunch{b_0}})) \mid \nu_b \rangle \sFun{\bunch{b}}{}{}$
$\langle Op: (\langle Add: () \mid \nu_e \rangle, (\nu \ \keyword{as} \ \langle Op: (\langle Add: () \mid \nu_e \rangle,(\nu, \nu)) \mid \nu_d \rangle, \nu \ \keyword{as} \ \langle Op: (\langle Add: () \mid \nu_e \rangle,(\nu, \nu)) \mid \nu_d \rangle)) \mid \nu_d \rangle_{\bunch{b_0}}$
\end{lstlisting}

By applying the conversion rule for equi-recursive types (which exist here because of type unifications), we get a simpler type where obviously irrelevant parts are grayed out.
\begin{lstlisting}[mathescape]
$\langle Op: (\langle Mul: ()_{\bunch{b}} \mid \textcolor{gray}{\nu_a} \rangle, (\langle 1: ()_{\bunch{b}} \mid \textcolor{gray}{2: () \mid \nu_c} \rangle, (\nu \ \keyword{as} \ \langle Op: (\langle Add: () \mid \nu_e \rangle,(\nu, \nu)) \mid \nu_d \rangle)_{\bunch{b_0}})) \mid \textcolor{gray}{\nu_b} \rangle \sFun{\bunch{b}}{}{}$
$(\nu \ \keyword{as} \ \langle Op: (\langle Add: () \mid \nu_e \rangle,(\nu, \nu)) \mid \nu_d \rangle)_{\bunch{b_0}}$
\end{lstlisting}

For this type, we can already see its connection with the expression \code{$\keyword{rule} \ Op \ (Mul \ (), (1 \ (), v)) \to v$}, and how the type representing a single rewrite path can be extracted with the assistance of the trace variables. When using syntactical sugar for numbers, binary operators and row variables, the presentation of the type becomes more clear:

\begin{lstlisting}[mathescape]
$\langle \langle 1_{\bunch{b}} \mid \textcolor{gray}{2} \mid \rangle \ \langle *_{\bunch{b}} \mid \rangle \ (\nu \ \keyword{as} \ \langle \nu \ \langle + \mid \rangle \ \nu \mid \rangle)_{\bunch{b_0}} \mid \rangle \sFun{\bunch{b}}{}{} (\nu \ \keyword{as} \ \langle \nu \ \langle + \mid \rangle \ \nu \mid \rangle)_{\bunch{b_0}}$
\end{lstlisting}

We can also see here the correspondence between the trace members and the pattern variables. Now we can drop the underlying structural types and the gray irrelevant parts and get a simplified type $\sFun{\bunch{b}}{1 * b_0}{b_0}$, which still clearly shows the rewriting performed by the corresponding rule. We can do the same for the other trace identified by $c$, and we will get another simplified type $\sFun{\bunch{c}}{2 * c_0}{c_0 + c_0}$. Putting these two types together and separating the parts identified by $b$ and $c$ with vertical bars, we finally get the informal type as originally shown in \Cref{sec:by-example}:
\begin{center}
  $\sFun{\bunch{b,c}}{ 1 * b_0 ~|~ 2 * c_0 }{ b_0 ~|~ c_0 + c_0 }$
\end{center}

\paragraph{Type schemes}
The type system of Typed \elevate is also equipped with generalized types. A type scheme (\(\tSigma\)) represents an universally quantified type, and carries a tracing environment $\Phi$, to be instantiated with fresh traces. Type schemes without tracing environment are denoted as \(\sigma\) and inductively defined in a standard way. The kind of the bound type variables are specified at the binding site. The $\many{\forall \ (\alpha_i: \kappa_i).}{,}{i}{0}{n}$ syntax may be used in the text to collectively describe a series of quantifiers when the order is insignificant.

\paragraph{Environments}
Figure \ref{fig:types} gives the definitions of various environments. The kinding (\(\Delta\)) and typing environments (\(\Gamma\)) are standard: they can be either empty (\(\cdot\)) or extended with a type variable and its kind (\(\nu: \kappa\)) or a variable and its type scheme (\(x: \tSigma\)). The tracing environment (\(\Phi\)) is a collection of traces, and it provides all the tracing information required by other environments and types. The following conversion rule is used if all the traces in the environment do not contain trace members.
\begin{center}
\code{$\many{\alpha_i \cstOf \bunch{}}{,}{i}{0}{n} = \varphi \cstOf \bunch{} \ if \ \varphi = \bunch{\many{\alpha_i}{,}{i}{0}{n}}$}
\end{center}
The rule environment (\(\Theta\)) is the typing environment used while defining a rewrite rule, and it contains the variable-type pairs extracted from the LHS of the rule. It is possible that a rule is defined without any variables on the LHS, so to clearly distinguish the inner and outer sides of rules, a rule environment can be inductively constructed from an empty environment ($\cdot$) or a nameless placeholders (\(\nop\)) which make the rule environment non-empty without introducing variables. The strategy combinator environment ($\Sigma$) is used while defining a strategy combinator, and it only holds variables of the strategy type. By the way, the $\mathbf{erase}$ function for erasing the traces in types can be easily generalized to be used on environments.

\paragraph{Linearity of Strategy Variables}
Our goal of designing this type system is to precisely reflect how strategies manipulate the rewritten programs, but some strategy definitions are very tricky to give a precise type. Let us consider, the challenging strategy combinator $\keyword{st} \ x \scFun \keyword{seq} \scApp x \scApp x$ that sequentially composes a strategy $x$ with itself. For assigning a type to this strategy combinator, we only know that $x$ has a general strategy type $\sFun{\varphi}{\tau_p}{\tau_e}$. As $x$ is sequentially composed with itself, its output type becomes its own input type, and therefore, its output type $\tau_e$ must match its input type $\tau_p$. Via unifications, the type of $x$ (and its sequential composition with itself) becomes the imprecise strategy type $\sFun{\varphi_{x}}{\tau_q}{\tau_q}$.
Following this procedure, we can arbitrarily extend the chain of sequential composition, for example, $\keyword{st} \ x \scFun \keyword{seq} \scApp (\keyword{seq} \scApp x \scApp x) \scApp x$, all resulting in the same imprecise type.
This is clearly in conflict with our goal of precise typing. To avoid this problem, we restrict the repeated use of strategy variables, such as $x$ in the example, with linearity to be used only once.
To enforce linearity we remove the contraction structural rule from the tracing environment and the strategy combinator environment, enforcing that there cannot be repeated occurrences of the same trace identifier or the same strategy variable.
More details about typing strategy combinators will be covered in \Cref{sec:st-typing}.
As a workaround to avoid this problem, we can simply bind a strategy to a name via \code{let} which allows the sequential composition of a strategy with itself in a safe way, as described in \Cref{sec:aux-typing}.

\paragraph{Typing Judgement}
Typing judgements are of the form $\Delta;\Gamma;\Phi^{\inc};\Sigma^{\inc};\Theta \vdash e:t \dashv \Phi^{\outc};\Sigma^{\outc}$, stating that the term $e$ has type $t$ under a series of input environments and the typing judgement produces environments as its output. The equivalence of types in this paper is up to alpha-renaming. If the name of an input environment collides with an output environment, the $\inc$ and $\outc$ superscripts are used to distinguish the input one and the output one, respectively. For the input environments on the left of the turnstile, $\Delta$ provides the kinding information, and it ensures that $t$ and the types in all other environments (including those output environments) are well-kinded; $\Gamma$ provides the type information for variables bound by $\keyword{let}$; $\Phi^{\inc}$ is the input tracing environment, and it ensures that the types in $\Sigma^{\inc}$ and $\Theta$ are well-traced; $\Sigma^{\inc}$ holds the input type of variables bound by $\keyword{st}$; $\Theta$ holds the type of variables bound by $\keyword{rule}$. For the output environments on the right of the reverse turnstile, $\Phi^{\outc}$ is the output tracing environment, and it ensures that $t$ and all types in $\Sigma^{\outc}$ are well-traced; $\Sigma^{\outc}$ holds the output type of variables bound by $\keyword{st}$. If any of the environment must be empty, we omit it from the typing judgement. For example, $\Delta;\Gamma;\Phi^{\inc};\Sigma^{\inc} \vdash e:t \dashv \Phi^{\outc};\Sigma^{\outc}$ is equivalent to $\Delta;\Gamma;\Phi^{\inc};\Sigma^{\inc};\cdot \vdash e:t \dashv \Phi^{\outc};\Sigma^{\outc}$. If the output environments are exactly the same as their input counterparts, we omit them together with the reverse turnstile. For example, $\Delta;\Gamma;\Phi^{\inc};\Sigma^{\inc};\Theta \vdash e:t$ is equivalent to $\Delta;\Gamma;\Phi^{\inc};\Sigma^{\inc};\Theta \vdash e:t \dashv \Phi^{\inc};\Sigma^{\inc}$.

\subsection{Well-formedness of Types}
\label{sec:well-formed}

\begin{figure}
  \footnotesize
  \begin{gather*}
  \frac{\nu:\kappa \in \Delta}{\Delta \vdash \nu:\kappa} \ \ruleName{K-Var}
  \quad
  \frac{}{\Delta \vdash ():\tdk} \ \ruleName{K-Unit}
  \quad
  \frac{\Delta \vdash \tau_m:\tdk \quad \Delta \vdash \tau_n:\tdk}{\Delta \vdash (\tau_m, \tau_n):\tdk} \ \ruleName{K-Pair}
  \\\\
  \frac{\Delta \vdash \rho:\rk}{\Delta \vdash \langle \rho \rangle:\tdk} \ \ruleName{K-Variant}
  \quad
  \frac{\Delta, \nu:\tdk \vdash \rho:\rk}{\Delta \vdash \nu \ \keyword{as} \ \langle \rho \rangle:\tdk} \ \ruleName{K-Recursive}
  \quad
  \frac{\Delta \vdash \tau: \tdk}{\Delta \vdash \tau:\tk} \ \ruleName{K-Subkind}
  \quad
  \frac{\Delta \vdash \tau: \tdk}{\Delta \vdash \tau_{\psi}:\tdk} \ \ruleName{K-Traced}
  \\\\
  \frac{\Delta \vdash \rho: \rk \quad \Delta \vdash \tau:\tdk \quad \ell \in \rk \quad \rk' = \rk \setminus \ell}{\Delta \vdash (\ell: \tau \mid \rho): \rk'} \ \ruleName{K-RowExt}
  \quad
  \frac{}{\Delta \vdash \cdot:\rk} \ \ruleName{K-EmptyRow}
  \\\\
  \frac{\Delta \vdash \tau_p:\tdk \quad \Delta \vdash \tau_e:\tdk \quad \Phi \vdash \tau_p \dashv \varphi \quad \Phi \vdash \tau_e \dashv \varphi}{\Delta;\Phi \vdash \sFun{\varphi}{\tau_p}{\tau_e}:\tk} \ \ruleName{K-Strategy}
  \quad
  \frac{\Delta \vdash \tau:\tdk \quad \Phi \vdash \tau \dashv \varphi}{\Delta;\Phi \vdash \sRes{\varphi}{\tau}:\tk} \ \ruleName{K-Result}
  \\\\
  \frac{\Delta;\Phi \vdash \sFun{\varphi}{\tau_p}{\tau_e}:\tk \quad \Delta;\Phi \vdash \omega:\tk}{\Delta;\Phi \vdash (\sFun{\varphi}{\tau_p}{\tau_e}) \scFun \omega:\tk} \ \ruleName{K-Combinator}
  \end{gather*}
\caption{Kinding rules}
\label{fig:kinding}
\end{figure}

\Cref{fig:kinding} shows the kinding rules checking the well-formedness of types. Kinding judgments are of the form $\Delta;\Phi \vdash t:\kappa$, stating that the type $t$ is well-formed and has kind $\kappa$ in the kinding environment $\Delta$ and the tracing environment $\Phi$. When the tracing environment is not used, we omit it from the judgement. Most of the kinding rules are straightforward. Most row kinding rules here are similar to those in \citep{daniel2016liberating}, except the $\ruleName{K-Variant}$ rule, which is more relaxed in our type system. Instead of requiring the row kind to be empty to construct a variant type, we do not have this requirement and the kind of $\rho$ in $\ruleName{K-Variant}$ can be arbitrary. This still ensures that there are no repeated labels in a row, and also gives us the flexibility of expressing the certain absence of some labels in a row as in \citep{blume06cases}. Apart from this, there are three specific kinding rules for our type system: $\ruleName{K-Strategy}$, $\ruleName{K-Result}$, and $\ruleName{K-Combinator}$. They are the kinding rules for traced types, checking for the well-formedness of both the traces and the underlying types. Thus, a set of tracing rules is required to define them.

\begin{figure}
  \footnotesize
  \begin{gather*}
  \frac{\Phi \vdash \tau_{\bunch{\gamma}} \dashv \varphi_h \quad \Phi \vdash ()_{\psi} \dashv \varphi_{t}}{\Phi \vdash \tau_{\bunch{\gamma, \psi}} \dashv \bunch{\varphi_{h}, \varphi_{t}}} \ \ruleName{Tr-Set}
  \quad
  \frac{\alpha \cstOf \bunch{\beta, \phi} \in \Phi \quad \Phi \vdash \tau \dashv \varphi_{\tau}}{\Phi \vdash \tau_{\bunch{\beta}} \dashv \bunch{\alpha, \varphi_{\tau}}} \ \ruleName{Tr-Member}
  \quad
  \frac{\alpha \cstOf \phi \in \Phi}{\Phi \vdash ()_{\bunch{\alpha}} \dashv \bunch{\alpha}} \ \ruleName{Tr-Identifier}
  \\\\
  \frac{}{\Phi \vdash () \dashv \bunch{}} \ \ruleName{Tr-Unit}
  \quad
  \frac{\Phi \vdash \tau_m \dashv \varphi \quad \Phi \vdash \tau_n \dashv \varphi}{\Phi \vdash (\tau_m, \tau_n) \dashv \varphi} \ \ruleName{Tr-Pair}
  \quad
  \frac{}{\Phi \vdash \nu \dashv \bunch{}} \ \ruleName{Tr-Variable}
  \\\\
  \frac{\Phi \vdash \tau \dashv \varphi_{\tau} \quad \Phi \vdash \langle \rho \rangle \dashv \varphi_{\rho}}{\Phi \vdash \langle \ell : \tau \mid \rho \rangle \dashv \bunch{\varphi_{\tau}, \varphi_{\rho}}} \ \ruleName{Tr-Variant}
  \quad
  \frac{}{\Phi \vdash \langle \cdot \rangle \dashv \bunch{}} \ \ruleName{Tr-EmptyRow}
  \\\\
  \frac{}{\Phi \vdash \langle \nu \rangle \dashv \bunch{}} \ \ruleName{Tr-RowVar}
  \quad
  \frac{\Phi \vdash \langle \rho \rangle \dashv \varphi}{\Phi \vdash \nu \ \keyword{as} \ \langle \rho \rangle \dashv \varphi} \ \ruleName{Tr-Recursive}
  \end{gather*}
\caption{Tracing rules}
\label{fig:well-traced}
\end{figure}

\Cref{fig:well-traced} shows the tracing rules which check the well-formedness of traces and also compute the trace identifier set for types. Tracing judgements are of the form $\Phi \vdash \tau \dashv \varphi$, stating that the type $\tau$ is well-traced in the tracing environment $\Phi$ and the corresponding trace identifier set for $\tau$ is $\varphi$. The most important tracing rule is $\ruleName{Tr-Set}$. For any traceable type $\tau$ traced by a trace variable set $\bunch{\gamma, \psi}$, we firstly check the well-formedness of $\tau$ traced by a single trace variable $\gamma$, which will be handled by either $\ruleName{Tr-Member}$ or $\ruleName{Tr-Identifier}$ depending on $\gamma$, and we get a trace identifier set $\varphi_h$, which identify $\gamma$ and the trace variables inside $\tau$. Since the internal traces in $\tau$ are already checked, we use a unit type to replace $\tau$ and compute the trace identifier set $\varphi_t$ of $()_{\psi}$, and then give the final result $\bunch{\varphi_h, \varphi_t}$. For the $\ruleName{Tr-Member}$ rule, we use trace members to trace the type of pattern variables which are terminals in the syntax. In other words, it is impossible for (the type of) a variable to have any internal (tracing) structure (identified by the same trace identifier). Thus, for $\tau_{\bunch{\beta}}$, $\beta$ must be identified by a variable $\alpha$ which is distinct from the elements in the trace identifier set of $\tau$. For the $\ruleName{Tr-Identifier}$ rule, the unit pattern and the unit type is the only non-variable terminal in the pattern and type syntax, respectively, so the unit type is the only type that can be directly traced by a trace identifier. Furthermore, there are tracing rules closely related with the structure of patterns (and the terms that can be matched by patterns). Both elements of a pair pattern must be present simultaneously, so the $\ruleName{Tr-Pair}$ require both elements of the pair type to have the same trace identifier set. Only one label can be present in a variant pattern, so the $\ruleName{Tr-Variant}$ rule require the trace identifier set of the type associated with each label $\ell$ to be distinct from the rest of the row. The rest of the tracing rules are straightforward.

\subsection{Typing Rules for Rewrite $\keyword{rule}$s}
\label{sec:rule-typing}

\begin{figure}
  \footnotesize
  \begin{gather*}
    \frac{
      \mathbf{fresh} \ \alpha \quad
      \Delta \vDash_{\alpha} p: \tau_p \dashv \Phi;\Theta \quad \Delta;\Gamma;\Phi;\Theta \vdash e : \tau_e
    }{
      \Delta;\Gamma;\Phi_{\Sigma};\Sigma \vdash \keyword{rule} \ p \to e : \sFun{\bunch{\alpha}}{\tau_p}{\tau_e} \dashv \Phi
    } \ \ruleName{T-R-Lam}
  \end{gather*}
  \vspace{.5em}\hrule\vspace{.5em}
  \begin{gather*}
  \frac{\mathbf{fresh} \ \beta \quad \Phi = \alpha \cstOf \bunch{\beta} \quad \Theta = \star, x: \tau_{\bunch{\beta}}}{\Delta \vDash_{\alpha} x : \tau_{\bunch{\beta}} \dashv \Phi;\Theta} \ \ruleName{P-Var}
  \quad
  \frac{\Phi = \alpha \cstOf \bunch{}}{\Delta \vDash_{\alpha} () : ()_{\bunch{\alpha}} \dashv \Phi;\star} \ \ruleName{P-Unit}
  \\\\
  \frac{
    \begin{gathered}
      \Phi = \alpha \cstOf \bunch{\phi_m, \phi_n} \quad \Theta_m \cap \Theta_n  = \emptyset\\
      \Delta \vDash_{\alpha} p_m:\tau_m \dashv (\alpha \cstOf \phi_m);\Theta_m \quad \Delta \vDash_{\alpha} p_n:\tau_n \dashv (\alpha \cstOf \phi_n);\Theta_n
    \end{gathered}
  }{\Delta \vDash_{\alpha} (p_m, p_n) : (\tau_m, \tau_n) \dashv \Phi;\Theta_m, \Theta_n} \ \ruleName{P-Pair}
  \quad
  \frac{\Delta \vDash_{\alpha} p:\tau \dashv \Phi;\Theta}{\Delta \vDash_{\alpha} \ell \ p : \langle \ell: \tau \mid \rho \rangle \dashv \Phi;\Theta} \ \ruleName{P-Label}
  \end{gather*}
  \vspace{.5em}\hrule\vspace{.5em}
  \begin{gather*}
  \frac{x:\tau \in \Theta}{\Delta;\Gamma;\Phi;\Theta \vdash x : \tau} \ \ruleName{T-R-Var}
  \quad
  \frac{\Phi = \alpha \cstOf \phi \quad \Theta \neq \cdot}{\Delta;\Gamma;\Phi;\Theta \vdash () : ()_{\bunch{\alpha}}} \ \ruleName{T-R-Unit}
  \\\\
  \frac{\Theta \neq \cdot \quad \Delta;\Gamma;\Phi;\Theta \vdash e_m:\tau_m \quad \Delta;\Gamma;\Phi;\Theta \vdash e_n:\tau_n}{\Delta;\Gamma;\Phi;\Theta \vdash (e_m, e_n) : (\tau_m, \tau_n)} \ \ruleName{T-R-Pair}
  \quad
  \frac{\Theta \neq \cdot \quad \Delta;\Gamma;\Phi;\Theta \vdash e:\tau}{\Delta;\Gamma;\Phi;\Theta \vdash \ell \ e : \langle \ell: \tau \mid \rho \rangle} \ \ruleName{T-R-Label}
  \end{gather*}
\caption{Typing rules for rewrite rules}
\label{fig:rule-typing}
\end{figure}

\Cref{fig:rule-typing} shows the typing rules for rewrite rules.

\paragraph{Typing Rule for Rewrite Rule Definitions}
We start with the central rule $\ruleName{T-R-Lam}$ for typing rewrite rule definitions. A fresh trace identifier $\alpha$ is assigned to this rule. A rule environment $\Theta$ (together with its tracing environment $\Phi$) is computed from the pattern $p$ or the LHS of the rewrite rule, and then $\Theta$ is used for typing the RHS of the rewrite rule $e$.

\paragraph{Pattern Typing Rules for the LHS}
The typing judgement for patterns is of the form $\Delta \vDash_{\alpha} p:\tau \dashv \Phi;\Theta$, stating that given the kinding environment $\Delta$ and the assigned trace identifier $\alpha$, the pattern $p$ has type $\tau$, and the judgement produces a tracing environment $\Phi$ and a rule environment $\Theta$. The only trace in the tracing environment $\Phi$ is identified by $\alpha$, and $\Phi$ ensures that $\tau$ and all types in $\Theta$ are well-traced. As for the pattern typing rules, $\ruleName{P-Var}$ assigns a fresh trace member $\beta$ to the underlying type $\tau$ of the pattern variable $x$, and then extend a non-empty rule environment with $x:\tau_{\bunch{\beta}}$ as its output; $\ruleName{P-Unit}$ simply takes the assigned trace identifier and produces a non-empty rule environment as its output; $\ruleName{P-Pair}$ ensures the pattern is linear by checking if the output environments from both elements do not intersect; $\ruleName{P-Label}$ gives a variant type containing the matched label $\ell$ and the corresponding type $\tau$, and according to the tracing rule, $\rho = \mathbf{erase}(\rho)$.

\paragraph{Typing Rules for RHS}
The remaining typing rules in \Cref{fig:rule-typing} are for typing the RHS of rewrite rules, and are straightforward. The condition $\Theta \neq \cdot$ is for prohibiting the usage of other language constructs in the body of rewrite rules. Noticeably, the typing rules for both sides of a rewrite rule definition are similar except the linear restriction for patterns, also sharing the same syntax. The reason for this is that the RHS of a rewrite rule can be matched against the LHS of another rewrite rule during compositions, and this similarity makes them comparable with each other.

\subsection{Typing Rules for $\keyword{st}$rategy Combinators}
\label{sec:st-typing}

\begin{figure}
\footnotesize
\begin{gather*}
\frac{\mathbf{fresh} \ \alpha, \beta_i, \beta_j, \beta_k \qquad \Phi = \alpha \cstOf \bunch{\beta_i, \beta_j, \beta_k}}{
  \Delta;\Gamma;\Phi_{\Sigma};\Sigma \vdash \keyword{seq}: (\sFun{\bunch{\alpha}}{\tau^i_{\bunch{\beta_i}}}{\tau^j_{\bunch{\beta_j}}}) \scFun (\sFun{\bunch{\alpha}}{\tau^j_{\bunch{\beta_j}}}{\tau^k_{\bunch{\beta_k}}}) \scFun (\sFun{\bunch{\alpha}}{\tau^i_{\bunch{\beta_i}}}{\tau^k_{\bunch{\beta_k}}}) \dashv \Phi
} \ \ruleName{T-Seq}
\\[1em]
\frac{\mathbf{fresh} \ \alpha_m, \alpha_j, \beta_m, \beta_n, \beta_j, \beta_k \qquad \Phi = \alpha_m \cstOf \bunch{\beta_m, \beta_n}, \alpha_j \cstOf \bunch{\beta_j, \beta_k}}{
  \Delta;\Gamma;\Phi_{\Sigma};\Sigma \vdash \keyword{choice}: (\sFun{\bunch{\alpha_m}}{\tau^p_{\bunch{\beta_m}}}{\tau^e_{\bunch{\beta_n}}}) \scFun (\sFun{\bunch{\alpha_j}}{\tau^p_{\bunch{\beta_j}}}{\tau^e_{\bunch{\beta_k}}}) \scFun (\sFun{\bunch{\alpha_m, \alpha_j}}{\tau^p_{\bunch{\beta_m, \beta_j}}}{\tau^e_{\bunch{\beta_n, \beta_k}}}) \dashv \Phi
} \ \ruleName{T-Choice}
\\[1em]
\frac{
  \begin{gathered}
  \mathbf{fresh} \ \alpha, \beta_m, \beta_n  \qquad \Phi^{\inc}_x = \alpha \cstOf \bunch{\beta_m, \beta_n} \qquad
  \omega_x = \sFun{\bunch{\alpha}}{\mathbf{erase}(\tau_m)_{\bunch{\beta_m}}}{\mathbf{erase}(\tau_n)_{\bunch{\beta_n}}}
  \\
  \Delta;\Gamma;\Phi^{\inc},\Phi^{\inc}_x;\Sigma^{\inc},x:\omega_x \vdash e_b : \omega_b \dashv \Phi^{\outc};\Sigma^{\outc}, x:\sFun{\varphi}{\tau_m}{\tau_n}
  \end{gathered}
}{
  \Delta;\Gamma;\Phi^{\inc};\Sigma^{\inc} \vdash \keyword{st} \ x \scFun e_b : (\sFun{\varphi}{\tau_m}{\tau_n}) \scFun \omega_b \dashv \Phi^{\outc};\Sigma^{\outc}
} \ \ruleName{T-S-Lam}
\\[1em]
\frac{\omega = \sFun{\bunch{\alpha}}{\tau_m}{\tau_n} \quad \alpha \cstOf \phi \in \Phi^{\inc} \quad x:\omega \in \Sigma^{\inc}}{\Delta;\Gamma;\Phi^{\inc};\Sigma^{\inc} \vdash x : \omega \dashv \alpha \cstOf \phi;x:\omega} \ \ruleName{T-S-Var}
\\[1em]
\frac{
  \begin{gathered}
  \Delta;\Gamma;\Phi^{\inc};\Sigma^{\inc} \vdash e_f: (\sFun{\varphi_f}{\tau_m}{\tau_n}) \scFun \omega_f \dashv \Phi^{\outc}_f;\Sigma^{\outc}_f
  \qquad
  \Delta;\Gamma;\Phi^{\inc};\Sigma^{\inc} \vdash e: \sFun{\varphi_e}{\tau_j}{\tau_k} \dashv \Phi^{\outc}_e;\Sigma^{\outc}_e
  \\
  \mathbf{erase}(\sFun{\varphi_f}{\tau_m}{\tau_n}) = \mathbf{erase}(\sFun{\varphi_e}{\tau_j}{\tau_k})
  \quad
  (\Phi^{\outc};\Sigma^{\outc};\omega) = \mathbf{compTrace}(\Phi^{\outc}_f;\Sigma^{\outc}_f;\sFun{\varphi_f}{\tau_m}{\tau_n};\omega_f;\ \Phi^{\outc}_e;\Sigma^{\outc}_e;\sFun{\varphi_e}{\tau_j}{\tau_k})
  \end{gathered}
}{\Delta;\Gamma;\Phi^{\inc};\Sigma^{\inc} \vdash e_f \scApp e: \omega \dashv \Phi^{\outc};\Sigma^{\outc}} \ \ruleName{T-S-App}
\end{gather*}
\caption{Typing rules for strategy combinators}
\label{fig:st-typing}
\end{figure}

\Cref{fig:st-typing} shows the typing rules for strategy combinators, including the typing rules for basic strategy combinators, strategy combinator definitions and strategy combinator applications.

\paragraph{Typing Rules for Basic Combinators}
The typing rules for the basic strategy combinators, $\keyword{seq}$ and $\keyword{choice}$, are straightforward. For $\keyword{seq}$, it sequentially composes two strategies, so it requires that the two strategies have the same trace and the output of the first strategy matches the input of the second one. For $\keyword{choice}$, it makes a nondeterministic choice between two strategies, so it requires that the two strategies have distinct traces, but their underlying structural type must be the same, and the two traces will be put together in the result type of the composition.

\paragraph{Typing Rules for Custom Strategy Combinators}
To construct strategy combinators, we introduced the $\keyword{st}$ keyword, and the corresponding typing rule is $\ruleName{T-S-Lam}$. According to the type rules for this type system, we distinguish the input and output environments, and only the output environments contain the final tracing information (after all trace computations). Thus, for the expression $\keyword{st} \ x \scFun e_b$, we have to extract the type of $x$ from the output strategy combinator environment to ensure that the type precisely reflect the usage of $x$ in $e_b$. Consequently, this requires a carefully designed type for $x$ in the input strategy combinator environment. In the $\ruleName{T-S-Lam}$ rule, $x$ is assigned an input type with a fresh trace identifier and two fresh trace members on the LHS and the RHS of the strategy type, respectively. Besides, the underlying type of $x$ always stay the same in the input and output environments, so it is directly used without any change. For this input type of $x$, an intuitive interpretation is that the type does not assume anything about tracing, and it has the most general trace, so when $x$ is (linearly) used and involved in trace computations, the output type will capture the tracing information from the context. Formally, the correctness of the $\ruleName{T-S-Lam}$ rule is justified by \Cref{lem:stSubst} which shows that beta-reductions for strategy combinators preserve types.

The rule $\ruleName{T-S-Var}$ is for the usage of the bound variable in the body of a strategy combinator definition. It takes the type of the variable $x$ from the input strategy combinator environment, and sets the output strategy combinator environment as a singleton only containing $x$ due to linearity.


\paragraph{Typing Rule for Strategy Combinator Application}
The $\ruleName{T-S-App}$ rule is one of the most complex typing rules in this type system.
It is for strategy combinator applications of the form $e_f \scApp e$, that applies the strategy $e$ to the strategy combinator $e_f$.
The rule calls $\mathbf{compTrace}$ computing the traced return type and environments by enumerating all possible ways of connecting each pair of traces in the type expected as a parameter by $e_f$ and the type of $e$ passed as an argument.
It then applies the computation result to the type of $e_f$'s body and the types in the strategy combinator environment.
Formally, $\ruleName{T-S-App}$ states that for a strategy combinator application $e_f \scApp e$ where $e_f$ has type $(\sFun{\varphi_f}{\tau_m}{\tau_n}) \scFun \omega_f$ and $e$ has type $\sFun{\varphi_e}{\tau_j}{\tau_k}$,
we start with checking if the underlying structural types of the parameter $e_f$ and the argument $e$ are the same, that is, checking the equality of their types after trace erasure.
Then we use the $\mathbf{compTrace}$ function to compute the type and the output environments of $e_f \scApp e$.
This process is much more involved than type-checking ordinary function applications because of the existence of traces.

\emph{Trace computation}
To understand how $\mathbf{compTrace}$ works, we look at its definition in \Cref{lst:compTrace},
which takes seven inputs that can be separated into two groups:
the first four arguments $(\Phi_m;\Sigma_m;\omega_m;\omega_c)$ are the output environments and traced types from the strategy combinator $e_f$, where $\omega_m$ is the parameter type and $\omega_c$ the type of the strategy combinator body;
the other three arguments $(\Phi_n;\Sigma_n;\omega_n)$ are the output environments and the traced type from the argument $e$.
%
Due to linearity, $\omega_m$ cannot share any common trace identifiers with $\omega_n$.
The trace computation is performed on $\omega_m$ and $\omega_n$, and the computation result is then applied to $\Sigma_m$, $\Sigma_n$, and $\omega_c$ to synthesize the resulting environments and traced type $(\Phi_r;\Sigma_r;\omega_r)$.
We use ($:=$) for mutable variable assignments and ($=$) for constant initialization.
When computing with traced types $\omega$, we must handle them together with their tracing and strategy combinator environments $\Phi$ and $\Sigma$.
Therefore, we call them together a \emph{traced triple} $(\Phi;\Sigma;\omega)$, where $\Sigma$ and $\omega$ is well-traced in $\Phi$.

\begin{lstlisting}[float=t, numbers=left, xleftmargin=2.2cm, xrightmargin=2.2cm, mathescape, caption={Definition of $\mathbf{compTrace}$}, captionpos=b, label={lst:compTrace}]
  $\mathbf{compTrace}(\Phi_m;\Sigma_m;\omega_m;\omega_c;\Phi_n;\Sigma_n;\omega_n) =$
    $\varphi_m = \mathbf{traceIds}(\Phi_m;\omega_m)$
    $\varphi_n = \mathbf{traceIds}(\Phi_n;\omega_n)$
    $(\Phi^{-}_m; \Sigma^{-}_m; \omega^{-}_c) = \mathbf{Delete} \ \varphi_m \ \mathbf{in} \ (\Phi_m; \Sigma_m; \omega_c)$
    $(\Phi^{-}_n; \Sigma^{-}_n; \omega^{-}_n) = \mathbf{Delete} \ \varphi_n \ \mathbf{in} \ (\Phi_n; \Sigma_n; \omega_n)$
    $(\Phi_r;\Sigma_r;\omega_r) := ((\Phi^{-}_m, \Phi^{-}_n);(\Sigma^{-}_m, \Sigma^{-}_n);\omega^{-}_c)$
    $\mathbf{for} \ \alpha_m \ \mathbf{in} \ \varphi_m$
      $\mathbf{for} \ \alpha_n \ \mathbf{in} \ \varphi_n$
        $(\alpha_m \cstOf \phi_m; \Sigma'_m; \omega'_m) = \mathbf{Select} \ \bunch{\alpha_m} \ \mathbf{in} \ (\Phi_m; \Sigma_m; \omega_m)$
        $(\alpha_m \cstOf \phi_m; \cdot; \omega'_c) = \mathbf{Select} \ \bunch{\alpha_m} \ \mathbf{in} \ (\Phi_m; \cdot; \omega_c)$
        $(\alpha_m \cstOf \phi_n; \Sigma'_n; \omega'_n) = (\mathbf{Select} \ \bunch{\alpha_n} \ \mathbf{in} \ (\Phi_n; \Sigma_n; \omega_n))[\alpha_n \mapsto \alpha_m]$
        $\overline{\mathcal{S}} = \mathbf{unifyTrace}(\alpha_m \cstOf \bunch{\phi_m, \phi_n};\omega'_m; \omega'_n)$
        $\mathbf{if} \ \overline{\mathcal{S}} \neq \mathbf{failTrace}$
          $(\Phi_i;\Sigma_i;\omega_i) = \mathbf{Fresh} \ (\alpha_m \cstOf \bunch{\phi_m, \phi_n}; (\Sigma'_m, \Sigma'_n)[\overline{\mathcal{S}}]; \omega'_c[\overline{\mathcal{S}}])$
          $(\Phi_r; \Sigma_r; \omega_r) := \mathbf{Add} \ (\Phi_i; \Sigma_i; \omega_i) \ \mathbf{to} \ (\Phi_r; \Sigma_r; \omega_r)$
    $\mathbf{return} \ (\Phi_r; \Sigma_r; \omega_r)$
\end{lstlisting}

The basic idea of $\mathbf{compTrace}$ is to enumerate all possible ways of connecting the traces in $\omega_m$ and $\omega_n$ and applying the result to $\omega_c$ and the strategy combinator environments. 
We start in lines 2--3 by extracting the trace identifier sets $\varphi_m$ and $\varphi_n$ from the traced triples for $\omega_m$ and $\omega_n$ using the $\mathbf{traceIds}$ function whose formal definition is in the appendix in \Cref{fig:traceIds}.
The extracted trace identifier sets contain the traces that are involved in the subsequent computations.
In lines 4--6, we remove the involved traces from the traced triple of $\omega_c$ and the traced triple of $\omega_n$ to obtain traced triples that only contain traces that are \emph{not} involved in the computation.
We combine the triples into $(\Phi_r;\Sigma_r;\omega_r)$, that is also the initial value of the accumulator for the subsequent loops.
The two nested loops in lines 7--8 enumerate all pairs of trace identifiers $(\alpha_m, \alpha_n)$ from $\varphi_m$ and $\varphi_n$.
In the loop body, we select the traced triples traced by $\alpha_m$ and $\alpha_n$ (lines 9--11) from the traced triples of the input types, to get single-traced slices.
We rename $\alpha_n$ to $\alpha_m$, in line 11, to ensure that all traces are temporarily uniformly identified and ready for subsequent computations.
The $\mathbf{Select}$ function, selects traces from the tracing environment and removes all not selected traces from the traced triple it returns. Its formal definition is in the appendix in \Cref{fig:select}.
$\mathbf{Select}$ is the dual of the $\mathbf{Delete}$ operator used before.
In line 12, we unify the traced types $\omega'_m$ and $\omega'_n$.
We will discuss trace unification in more depth below.
If the unification succeeded (line 13) with a result $\overline{\mathcal{S}}$, there is a way to connect the traces identified by $\alpha_m$ and $\alpha_n$, meaning that there exists a possible execution path for the strategy.
Subsequently, in line 14, we apply $\overline{\mathcal{S}}$ to a traced triple containing the strategy combinator environments ($\Sigma'_m, \Sigma'_n$) which can be safely put together because of linearity, and the single-traced return type of $e_f$ ($\omega'_c$). We $\mathbf{Fresh}$ly rename the traced triples before $\mathbf{Add}$ing it $\mathbf{to}$ the accumulator in line 15 to enforce the distinction between trace identifiers. Finally, the accumulator is returned as the result of $\mathbf{compTrace}$ (line 16).
The formal definitions of $\mathbf{Fresh}$ and $\mathbf{Add}$ can be found in the appendix in \Cref{fig:fresh} and \Cref{fig:add}.

\begin{figure}
\footnotesize
\begin{gather*}
  \begin{aligned}
    &\mathbf{unifyTrace}: (\Phi;t;t) \to \overline{\mathcal{S}}\\
    &\mathbf{unifyTrace}(\alpha \cstOf \phi;()_{\bunch{\alpha}};()) = \mathbf{failTrace} \\
    &\mathbf{unifyTrace}(\alpha \cstOf \phi;();()_{\bunch{\alpha}}) = \mathbf{failTrace} \\
    &\mathbf{unifyTrace}(\alpha \cstOf \bunch{\beta,\phi};\tau_{\bunch{\beta}};\tau_n) = \mathbf{if} \ \tau_n = \tau \ \mathbf{then} \ \mathbf{failTrace} \ \mathbf{else} \ \tau_{\bunch{\beta}} \mapsto \tau_n \\
    &\mathbf{unifyTrace}(\alpha \cstOf \bunch{\beta,\phi};\tau_m;\tau_{\bunch{\beta}}) = \mathbf{if} \ \tau_m = \tau \ \mathbf{then} \ \mathbf{failTrace} \ \mathbf{else} \ \tau_{\bunch{\beta}} \mapsto \tau_m \\[-.25em]
    &\cdots\\[-.25em]
    &\mathbf{unifyTrace}(\Phi;\tau;\tau) = (\cdot \mapsto \cdot) \\[-.25em]
    &\cdots\\[-.25em]
    &\mathbf{unifyTrace}(\Phi;\sRes{\bunch{\alpha}}{\tau_m};\sRes{\bunch{\alpha}}{\tau_n}) = \mathbf{unifyTrace}(\Phi;\tau_m;\tau_n) \\
    &\mathbf{unifyTrace}(\Phi;\sFun{\bunch{\alpha}}{\tau_m}{\tau_n}; \sFun{\bunch{\alpha}}{\tau_j}{\tau_k}) = \mathbf{let} \ \overline{\mathcal{S}} = \mathbf{unifyTrace}(\Phi;\tau_m; \tau_j) \\
    &\quad \mathbf{in} \ \mathbf{unifyTrace}(\Phi;\tau_n[\overline{\mathcal{S}}]; \tau_k[\overline{\mathcal{S}}]) \circ \overline{\mathcal{S}}
  \end{aligned}
\end{gather*}
\caption{Definition of trace unification}
\label{fig:trace-uni-part}
\end{figure}

\paragraph{Trace Unification}
Like ordinary type unifications, for traced type $\omega_m$ and $\omega_n$, trace unification produces a trace substitution $\overline{\mathcal{S}}$ such that $\omega_m[\overline{\mathcal{S}}] = \omega_n[\overline{\mathcal{S}}]$.
Trace substitutions are collections of mappings, but when a mapping in a ordinary substitution replaces a variable with a term, each mapping in a trace substitution replaces a traceable type traced by a trace member with another traceable type, that is $\tau_{\bunch{\beta}}[\overline{\mathcal{S}}] = \overline{\mathcal{S}}(\tau_{\bunch{\beta}})$ if $\tau_{\bunch{\beta}} \in dom(\overline{\mathcal{S}})$. Actually, there is no need to check $\tau$ while looking up $\tau_{\bunch{\beta}}$ in $dom(\overline{\mathcal{S}})$ because $\beta$ is already the unique key for the corresponding mapping (see \Cref{lem:unique}). Furthermore, trace substitutions are always generated by $\mathbf{unifyTrace}$ in this type system, and $\mathbf{unifyTrace}$ can only be performed on types whose underlying types are the same, so we have $\tau = \mathbf{erase}(\overline{\mathcal{S}}(\tau_{\bunch{\beta}}))$. In other words, trace substitutions never affect the underlying types. Trace substitutions naturally adapt the operations for ordinary substitutions, but for a concise formalization, we extend trace substitutions with a bottom value $\mathbf{failTrace}$ to represent the failure of $\mathbf{unifyTrace}$, and the compositions of any trace substitutions and $\mathbf{failTrace}$ will result in $\mathbf{failTrace}$.

A definition of the function $\mathbf{unifyTrace}$ is shown in \Cref{fig:trace-uni-part} focusing on the most important parts, and a complete definition can be found in the appendix in \Cref{fig:trace-uni}.
The definition is straightforward and very close to a standard type-unification definition.
$\mathbf{failTrace}$ is returned when we try to unify a traced type with its untraced counterpart, otherwise a trace substitution is returned. It is worth noting that $\mathbf{unifyTrace}$ is not defined for and never applied to strategy combinator types. \Cref{lem:uni} shows that $\mathbf{unifyTrace}$ is a correctly defined trace unification.

\begin{lemma}[Unique Underlying Type]
\label{lem:unique}
If $\Delta \vdash e_m:\omega \dashv \Phi$, for each $\alpha \cstOf \phi \in \Phi$, there is $(\alpha \cstOf \phi;\cdot;\omega_m) = \mathbf{Select} \ \bunch{\alpha} \ \mathbf{in} \ (\Phi; \cdot; \omega)$, and for each $\beta \in \phi$, it always traces an unique $\tau$ in $\omega_m$.
\end{lemma}
\begin{proof}
By induction on the typing derivation of $\Delta \vdash e_m:\omega \dashv \Phi$.
\end{proof}

\begin{lemma}[Unification]
\label{lem:uni}
If $\mathbf{unifyTrace}(\Phi;\omega_1;\omega_2) = \overline{\mathcal{S}} \neq \mathbf{failTrace}$, then $\omega_1[\overline{\mathcal{S}}] = \omega_2[\overline{\mathcal{S}}]$.
\end{lemma}
\begin{proof}
By induction on $\omega_1$ and the definition of $\mathbf{unifyTrace}$ and \Cref{lem:unique}.
\end{proof}

\begin{figure}[t]
  \includegraphics[width=\textwidth]{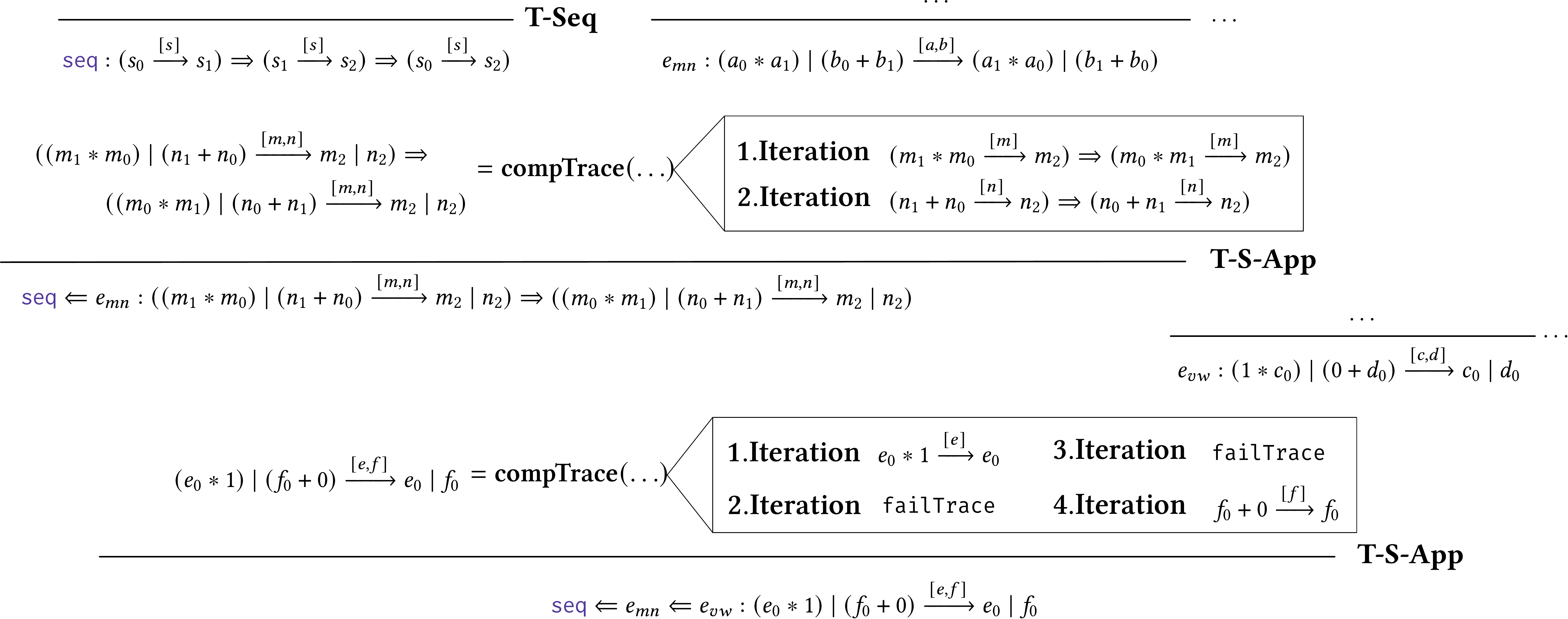}
  \caption{Simplified typing derivation for \emph{Example 3}}
  \label{fig:example_t_s_app}
\end{figure}

\paragraph{Example of Typing Strategies}
\Cref{fig:example_t_s_app} shows a simplified typing derivation to demonstrate the typing of the sequential composition from Example 3.
In the figure, we write $e_{mn}$ and $e_{vw}$ for the two composed strategies and use the simplified types for visual clarity.

To type the expression $\keyword{seq} \scApp e_{mn} \scApp e_{vw}$, we type $\keyword{seq} \scApp e_{mn}$ and $e_{vw}$ separately and then compute the resulting traced type based on their types.
Let's look at the typing of $\keyword{seq} \scApp e_{mn}$ in the top-left of the figure more closely.
Here, we type $\keyword{seq}$ and $e_{mn}$ at the very top and use their types as input to the $\mathbf{compTrace}$ function.
In this instance, there are two iterations as we connect one trace from $\keyword{seq}$ ($s$) with two traces from $e_{mn}$ ($a$ and $b$).
Each iteration performs trace unification resulting here in two traced types that are combined into the resulting traced type of the strategy application.
For the second strategy application in the bottom half of the figure, we compute the traces of two strategies each with two traces, therefore, $\mathbf{compTrace}$ performs $4\, (= 2 \times 2)$ iterations enumerating all possible ways to connect a trace from the first type with one from the second.
Here, for not all iterations the trace unification is successful, as not all traces connect.
The two successfully unified traced types are combined, resulting in the final traced type of the overall strategy application.

\subsection{Typing Rules for Strategy Executions and $\keyword{let}$-bindings}
\label{sec:aux-typing}

\begin{figure}[b]
  \footnotesize
  \begin{gather*}
\frac{}{\Delta;\Gamma \vdash \keyword{fail}: \sRes{\varphi}{\tau} \dashv \varphi \cstOf \bunch{}} \ \ruleName{T-Fail}
\\[.9em]
\frac{
  \begin{gathered}
  \mathbf{fresh} \ \alpha \quad \Delta;\Gamma;\alpha \cstOf \bunch{};\star \vdash e: \tau_e \qquad 
  \mathbf{erase}(\tau_a) = \mathbf{erase}(\tau_e)\\
  (\varphi_{s} \cstOf \bunch{}; \cdot; \sRes{\varphi_{s}}{\tau_{s}}) = \mathbf{Add} \ (\varphi_{a} \cstOf \bunch{}; \cdot; \sRes{\varphi_{a}}{\tau_{a}}) \ \mathbf{to} \ (\alpha \cstOf \bunch{}; \cdot; \sRes{\bunch{\alpha}}{\tau_e})
  \end{gathered}}{\Delta;\Gamma \vdash \keyword{succ} \ e: \sRes{\varphi_{s}}{\tau_{s}} \dashv \varphi_{s} \cstOf \bunch{}} \ \ruleName{T-Succ}
\\[.9em]
\frac{
  \begin{aligned}
  &\Delta;\Gamma \vdash e_a: \sRes{\varphi}{\tau} \dashv \varphi \cstOf \bunch{} \quad \Delta;\Gamma \vdash e_b: \sRes{\varphi}{\tau} \dashv \varphi \cstOf \bunch{}
  \end{aligned}
  }{\Delta;\Gamma \vdash e_a \alt e_b : \sRes{\varphi}{\tau} \dashv \varphi \cstOf \bunch{}} \ \ruleName{T-Alt}
\\[.8em]
\frac{
  \begin{gathered}
  \Delta;\Gamma \vdash e_f: \sFun{\varphi_f}{\tau_p}{\tau_e} \dashv \Phi_f \quad \Delta;\Gamma \vdash e_i: \sRes{\varphi_{i}}{\tau_{i}} \dashv \varphi_i \cstOf \bunch{}\qquad
  \mathbf{erase}(\tau_p) = \mathbf{erase}(\tau_i) \quad \mathbf{erase}(\tau_a) = \mathbf{erase}(\tau_e)\\
  (\varphi_r \cstOf \bunch{};\cdot;\sRes{\varphi_r}{\tau_r}) = \mathbf{Add} \ (\varphi_a \cstOf \bunch{}; \cdot; \sRes{\varphi_a}{\tau_a}) \ \mathbf{to} \ \mathbf{compTrace}(\Phi_f;\cdot;\sRes{\varphi_f}{\tau_p};\sRes{\varphi_f}{\tau_e};\varphi_i \cstOf \bunch{};\cdot;\sRes{\varphi_{i}}{\tau_{i}})
  \end{gathered}
  }{\Delta;\Gamma \vdash e_f \gets e_i : \sRes{\varphi_r}{\tau_r} \dashv \varphi_r \cstOf \bunch{}} \ \ruleName{T-Exec}
  \end{gather*}
\caption{Typing rules for strategy execution and results}
\label{fig:exec-typing}
\end{figure}

\paragraph{Typing Rules for Strategy Execution}
Being able to define rewrite rules and compose them with strategy combinators, we still need to execute the strategies and get the results. \Cref{fig:exec-typing} shows the typing rules for strategy execution and results. Executing a strategy gives two possible outcomes, $\keyword{fail}$ or $\keyword{succ} \ e$, and the execution is expected to preserve types. Thus, the typing rule $\ruleName{T-Fail}$ allows the $\keyword{fail}$ result to have an arbitrary underlying type and arbitrary traces because we know nothing about the strategy execution returning $\keyword{fail}$. It is worth noting that the output tracing environment does not contain any trace members because there cannot be any pattern variables in a closed term. Actually, all the expressions of the strategy result type are closed with respect to the rule environments and strategy combinator environments because in practice we always apply strategies to ASTs which are closed terms in the strategy language.
In the rule $\ruleName{T-Succ}$, if $\keyword{succ} \ e$ is the execution result of a strategy, the expression $e$ must have been constructed by the RHS of a rewrite, so there must be a trace for $e$ and $e$ is typed in the same way as the RHS of a rewrite rule.
Similarly, we also allow arbitrary traces to be added for type preservation. The $\ruleName{T-Alt}$ rule is for the non-deterministic choice operator $\alt$ which choose one from two operands. It is intuitive to require the two operands to have the same type. Finally, the relatively complex rule $\ruleName{T-Exec}$ is for strategy executions. The structure and mechanism of this typing rule is almost the same as those of the $\ruleName{T-S-App}$ rule for strategy combinator applications, and the $\mathbf{compTrace}$ function is also used to compute the traces in the execution results. As before, arbitrary traces can be added to the final type for type preservation.

\begin{figure}
  \footnotesize
  \begin{gather*}
  \frac{
    \Delta, \many{\alpha_i:\kappa_i}{,}{i}{0}{n};\Gamma \vdash e_f:\omega_f \dashv \Phi_f \qquad \Phi_f \neq \cdot
    \qquad
    \Delta;\Gamma, x:\keyword{tr} \ \Phi_f. \many{\forall \ (\alpha_i: \kappa_i).}{,}{i}{0}{n} \ \omega_f \vdash e: \omega_e \dashv \Phi_e
  }{\Delta;\Gamma \vdash \keyword{let} \ x = e_f \ \keyword{in} \ e: \omega_e \dashv \Phi_e} \ \ruleName{T-Let}
\\[.9em]
  \frac{
    x:\keyword{tr} \ \Phi_x. \ \sigma_x \in \Gamma \qquad
    \Delta \vdash \sigma_x \instRel \omega_x \quad (\Phi;\cdot;\omega) = \mathbf{Fresh} \ (\Phi_x; \cdot; \omega_x)
  }{\Delta;\Gamma;\Phi_{\Sigma};\Sigma \vdash x: \omega \dashv \Phi} \ \ruleName{T-Var}
\\[.9em]
  \frac{\Delta \vdash \rho: \rk' \quad \rk \subseteq \rk' \quad \rho = \mathbf{erase}(\rho) \quad \Delta \vdash \sigma[\nu \mapsto \rho] \instRel \omega}{\Delta \vdash \forall \ (\nu:\rk). \ \sigma \instRel \omega} \ \ruleName{I-Row}
\\[.9em]
  \frac{\Delta \vdash \tau: \tdk \quad \tau = \mathbf{erase}(\tau) \quad \Delta \vdash \sigma[\nu \mapsto \tau] \instRel \omega}{\Delta \vdash \forall \ (\nu:\tdk). \ \sigma \instRel \omega} \ \ruleName{I-Type}
  \end{gather*}
\caption{Typing rules for let bindings}
\label{fig:let-typing}
\end{figure}

\paragraph{Typing Rules for Let Bindings}
The last typing rules are shown in \Cref{fig:let-typing}, including the rules for $\keyword{let}$-binding, the usage of $\keyword{let}$-bound variables, and the type variable instantiation rules. All these rules are straightforward. Considering that the abstractions of traces and the underlying type variables are syntactically separated, the operations on the underlying types in these typing rules are also standard. Since our goal is to let traces precisely reflect the internal rewriting paths of strategies, the abstractions and instantiations of traces should preserve as much tracing information as possible while avoiding violating the sub-structural rules of the tracing environments. Thus, in the $\ruleName{T-Let}$ rule, the output tracing environment $\Phi_f$ of $e_f$ is directly bound by the $\keyword{tr}$ keyword in the type scheme, and in the $\ruleName{T-Var}$ rule, the bound tracing environment is freshly renamed (together with the traces in the traced type). By alpha-equivalence, the tracing information does not change, and there will not be any name collisions because all trace variables are freshly renamed. Besides, the $\ruleName{T-Let}$ rule will trigger the error in Example 6 and Example 7 by requiring that $\Phi_f$ is non-empty. The type variable instantiation rules are of the form $\Delta \dashv \sigma \instRel \omega$, stating that the $\sigma$ is a type scheme without a tracing environment, and it can be instantiated to a traced type $\omega$ in the kinding environment $\Delta$. The instantiations for rows and types are standard, except that the types or rows the variables are instantiated to must not contain any trace variables, so we can clearly distinguish the trace and type variable abstractions and preserve the tracing information.

\paragraph{Workaround for Linearity of Strategy Variables}
Recalling the linear restriction mentioned in \Cref{sec:type-syntax}, we forbid the definition of strategy combinators containing repeated usage of strategy variables such as $\keyword{st} \ x \scFun \keyword{seq} \scApp x \scApp x$. A workaround for this restriction is that we can define the strategy in a $\keyword{let}$-binding and repeatedly use the polymorphic $\keyword{let}$-bound variable, $f$ for example, because for each time $f$ is used, the bound tracing environment will be freshly renamed and there will not be common trace variables in the types of all occurrences of $f$.

\subsection{Operational Semantics}
\label{sec:op-semantics}

\begin{figure}
  \footnotesize
    \begin{alignat*}{4}
      \mathrm{Combinator \ Values} \quad
      &cv \ &\metaDef \quad & \keyword{seq} \cmid \keyword{choice} \cmid cv \scApp v \\
      \mathrm{Values} \quad
      &v \ &\metaDef \quad & cv \cmid \keyword{st} \ x \scFun e_b \cmid \keyword{rule} \ p \to e \cmid 
      \ell \ v \cmid () \cmid (v_m, v_n) \cmid \keyword{succ} \ v \cmid \keyword{fail}\\
      \mathrm{Evaluation \ Contexts} \quad
      &E \ &\metaDef \quad & [] \cmid E \scApp e \cmid v \scApp E \cmid E \gets e \cmid v \gets E \cmid \\
      &&&\ell \ E \cmid (E, e) \cmid (v, E) \cmid \keyword{let} \ x = E \ \keyword{in} \ e \cmid \keyword{succ} \ E \cmid E \alt e \cmid v \alt E
    \end{alignat*}
  \caption{Syntax of values and evaluation contexts.}
  \label{fig:value}
\end{figure}

\Cref{fig:value} shows the syntax of values (denoted as $v$) and evaluation contexts (denoted as $E$) defined in the style used in \citep{wright94soundness}. In Typed \elevate, primitive strategy combinators and their applications to values are considered as a subset of values, the combinator values (denoted as $cv$). From this perspective, primitive strategy combinators are similar to the data constructors in Haskell. The rest of the value and evaluation context definitions are straightforward.

\begin{figure}
  \footnotesize
  \begin{minipage}{.63\textwidth}
    \begin{gather*}
    \begin{aligned}
      E[e_a] &\evalRel E[e_b] \ \text{iff} \ e_a \stepRel e_b
      \\
      (v \gets \keyword{fail}) &\stepRel \keyword{fail}
      \\
      ((\keyword{seq} \scApp v_a \scApp v_b) \gets v_i) &\stepRel (v_b \gets (v_a \gets v_i))
      \\
      ((\keyword{choice} \scApp v_a \scApp v_b) \gets v_i) &\stepRel (v_a \gets v_i) \alt (v_b \gets v_i)
      \\
      ((\keyword{rule} \ p \to e) \gets \keyword{succ} \ v) &\stepRel \keyword{fail} \ \text{if} \ \mathbf{match}(p;v) = \mathbf{failPat}
      \\
      ((\keyword{rule} \ p \to e) \gets \keyword{succ} \ v) &\stepRel \keyword{succ} \ e[\mathcal{S}] \ \text{if} \ \mathbf{match}(p;v) = \mathcal{S} \neq \mathbf{failPat}
      \\
      \keyword{fail} \alt \keyword{succ} \ v_b &\stepRel \keyword{succ} \ v_b
      \\
      \keyword{succ} \ v_a \alt \keyword{succ} \ v_b &\stepRel \keyword{succ} \ v_a
      \\
      \keyword{succ} \ v_a \alt \keyword{succ} \ v_b &\stepRel \keyword{succ} \ v_b
      \\
      \keyword{succ} \ v_a \alt \keyword{fail} &\stepRel \keyword{succ} \ v_a
      \\
      \keyword{fail} \alt \keyword{fail} &\stepRel \keyword{fail}
      \\
      (\keyword{st} \ x \scFun e) \scApp v &\stepRel e[x \mapsto v]
      \\
      \keyword{let} \ x = v \ \keyword{in} \ e &\stepRel e[x \mapsto v]
      \end{aligned}
    \end{gather*}
  \end{minipage}%
  \hfill\vline\hfill%
  \begin{minipage}{.32\textwidth}
    \begin{gather*}
    \begin{aligned}
      &\mathbf{match}: p \to v \to \mathcal{S} \\
      &\mathbf{match}(x;v) = x \mapsto v \\
      &\mathbf{match}(\ell \ p;\ell \ v) = \mathbf{match}(p;v) \\
      &\mathbf{match}(();()) = (\cdot \mapsto \cdot) \\
      &\mathbf{match}((p_a, p_b);(v_a, v_b)) = \\
      &\qquad \mathbf{match}(p_b;v_b) \circ \mathbf{match}(p_a;v_a) \\
      &\mathbf{match}(p;v) = \mathbf{failPat}
    \end{aligned}
    \end{gather*}
  \end{minipage}
\caption{Operational Semantics}
\label{fig:op-semantics}
\end{figure}

\Cref{fig:op-semantics} shows the reduction rules in the operational semantics of Typed \elevate and the definition of a function $\mathbf{match}$ performing pattern matching. The relation $e_m \stepRel e_n$ contains the notions of reduction which are the interesting cases of $e_m$ reducing to $e_n$, and by embedding this relation into the evaluation contexts, we get the stepping relation $e_m \evalRel e_n$ which works for any redex \cite{wright94soundness}. All these definitions are straightforward. Being similar to $\mathbf{unifyTrace}$, the substitutions produced by $\mathbf{match}$ are also extended with a bottom value $\mathbf{failPat}$ to represent failed pattern matching. \Cref{lem:pat} shows that the $\mathbf{match}$ function is correctly defined.

\begin{lemma}[Pattern Matching]
\label{lem:pat}
If $\mathbf{match}(p;v) = \mathcal{S} \neq \mathbf{failPat}$, then $\mathbf{p2e}(p)[\mathcal{S}] = v$.
\end{lemma}
\begin{proof}
By the definition of $\mathbf{match}$ and induction on $p$.
\end{proof}

It is worth noting that the introduction of combinator values is a design choice we have made. As shown in the \Cref{fig:exec-typing}, we compromise on the precision of traces for type preservation during the execution of strategies. In the meantime, as shown by the reduction rules for $\keyword{seq}$ and $\keyword{choice}$, the primitive strategy combinators are interpreted as the compositions of strategy executions. Thus, to contain the compromise on precision in a limited range, we always organize the compositions of strategies as combinator values with precise traces before executing. A positive side effect of this design choice is that it also makes the extension of primitive strategy combinators easier.

\subsection{Soundness and Strong Normalization}
\label{sec:type-prop}

In this section, we show the soundness of our tracing type system, and its strong normalization property. In this and the subsequent sections, for a relation $\mathcal{R}$, we write $\mathcal{R}^{+}$ for its transitive closure, and $\mathcal{R}^{\ast}$ for its reflexive transitive closure.

Before presenting the type soundness theorem, we show below a series of substitution lemmata. \Cref{lem:rlSubst} and \Cref{lem:rlFail} are for the executions of rewrite rules. \Cref{lem:rlSubst} states that if the pattern matching in a non-empty traced rewrite rule execution succeeds, the execution is type-preserving, while \Cref{lem:rlFail} states that an empty-traced rewrite rule execution must fail.

\begin{lemmaE}[Rule Substitution]
\label{lem:rlSubst}
If $\Delta;\Gamma \vdash (\keyword{rule} \ p \to e) \gets \keyword{succ} \ v : \sRes{\bunch{\alpha_r}}{\tau_r} \dashv \alpha_r \cstOf \bunch{}$, and $\mathbf{match}(p;v) = \mathcal{S} \neq \mathbf{failPat}$, then $\Delta;\Gamma \vdash \keyword{succ} \ e[\mathcal{S}]: \sRes{\bunch{\alpha_r}}{\tau_r} \dashv \alpha_r \cstOf \bunch{}$.
\end{lemmaE}
\begin{proof}
The proof proceeds by induction on the derivation of the typing judgement for $p$.
\begin{proofEnd}
By inversion on $\ruleName{T-Exec}$ and $\ruleName{T-R-Lam}$, we know $\Delta \vDash_{\alpha_f} p: \tau_p \dashv \Phi;\Theta$, and $\Delta;\Gamma;\Phi;\Theta \vdash e : \tau_e$, and $\Delta;\Gamma;\alpha_v \cstOf \bunch{};\star \vdash v: \tau_v$, and $\mathbf{erase}(\tau_p) = \mathbf{erase}(\tau_v)$, and there is a $\overline{\mathcal{S}}$ generated by $\mathbf{unifyTrace}(\alpha_f \cstOf \phi_f;\sRes{\bunch{\alpha_f}}{\tau_p};\sRes{\bunch{\alpha_f}}{(\tau_v[\alpha_v \mapsto \alpha_f])})$. By induction on the typing derivation of $\Delta \vDash_{\alpha_f} p: \tau_p \dashv \Phi;\Theta$, and with $\mathbf{match}(p;v) \neq \mathbf{failPat}$ and the definition of $\mathbf{unifyTrace}$, we know that $\overline{\mathcal{S}} \neq \mathbf{failTrace}$. Then by \Cref{lem:uni} and \Cref{lem:pat}, we know that $\tau_v[\alpha_v \mapsto \alpha_f] = \tau_p[\overline{\mathcal{S}}]$, and for each variable $x$ in $p$ and also in $dom(\mathcal{S})$, there is a corresponding $\tau_{\bunch{\beta}}$ in $\tau_p$ and also in $dom(\overline{\mathcal{S}})$, and $\Delta;\Gamma;\alpha_f \cstOf \bunch{};\star \vdash \mathcal{S}(x): \overline{\mathcal{S}}(\tau_{\bunch{\beta}})$. Then it is easy to show that $\Delta;\Gamma;\alpha_f \cstOf \bunch{};\star \vdash e[\mathcal{S}]: \tau_e[\overline{\mathcal{S}}]$, which with appropriate renaming provides evidence for the proof goal.
\end{proofEnd}
\end{proof}

\begin{lemmaE}[Failed Rule]
\label{lem:rlFail}
If $\Delta;\Gamma \vdash (\keyword{rule} \ p \to e) \gets \keyword{succ} \ v : \sRes{\bunch{}}{\tau_r}$, then $(\keyword{rule} \ p \to e) \gets \keyword{succ} \ v \stepRel \keyword{fail}$.
\end{lemmaE}
\begin{proof}
The proof proceeds by induction on the derivation of the typing judgement for $p$.
\begin{proofEnd}
By inversion on $\ruleName{T-Exec}$ and $\ruleName{T-R-Lam}$, we know $\mathbf{failTrace} = \mathbf{unifyTrace}(\alpha_f \cstOf \phi_f;\sRes{\bunch{\alpha_f}}{\tau_p};\sRes{\bunch{\alpha_f}}{(\tau_v[\alpha_v \mapsto \alpha_f])})$, where $\Delta \vDash_{\alpha_f} p: \tau_p \dashv \Phi;\Theta$, and $\Delta;\Gamma;\Phi;\Theta \vdash e : \tau_e$, and $\Delta;\Gamma;\alpha_v \cstOf \bunch{};\star \vdash v: \tau_v$, and $\mathbf{erase}(\tau_p) = \mathbf{erase}(\tau_v)$. By induction on the typing derivation of $\Delta \vDash_{\alpha_f} p: \tau_p \dashv \Phi;\Theta$ and the definition of $\mathbf{unifyTrace}$, we know that the only source of $\mathbf{failTrace}$ is mismatched labels in $p$ and $v$. By the definition of $\mathbf{match}$, we know the result of $\mathbf{match}(p;v)$ must be $\mathbf{failPat}$. Thus, $(\keyword{rule} \ p \to e) \gets \keyword{succ} \ v \stepRel \keyword{fail}$.
\end{proofEnd}
\end{proof}

\Cref{lem:stSubst} shows that the beta-reductions of strategy combinator applications are type-preserving. It also justifies the design of the typing rule $\ruleName{T-S-Lam}$.

\begin{lemmaE}[Strategy Substitution]
\label{lem:stSubst}
If $\Delta;\Gamma;\Phi^{\inc};\Sigma^{\inc} \vdash (\keyword{st} \ x \scFun e_f) \scApp e : \omega \dashv \Phi^{\outc};\Sigma^{\outc}$, then $\Delta;\Gamma;\Phi^{\inc};\Sigma^{\inc} \vdash e_f[x \mapsto e] : \omega \dashv \Phi^{\outc};\Sigma^{\outc}$.
\end{lemmaE}
\begin{proof}
The proof proceeds by induction on the derivation of the typing judgement for $e_f$.
\begin{proofEnd}
By inversion on $\ruleName{T-S-App}$ and $\ruleName{T-S-Lam}$, we know that:
\\
$\Delta;\Gamma;\Phi^{\inc},\Phi^{\inc}_x;\Sigma^{\inc},x:\omega^{\inc}_x \vdash e_f : \omega_f \dashv \Phi_f^{\outc};\Sigma_f^{\outc}, x:\omega^{\outc}_x \quad \Delta;\Gamma;\Phi^{\inc};\Sigma^{\inc} \vdash e: \omega_e \dashv \Phi^{\outc}_e;\Sigma^{\outc}_e$
\\
$\mathbf{erase}(\omega^{\outc}_x) = \mathbf{erase}(\omega_e) \quad (\Phi^{\outc};\Sigma^{\outc};\omega) = \mathbf{compTrace}(\Phi^{\outc}_f;\Sigma^{\outc}_f;\omega^{\outc}_x;\omega_f;\Phi^{\outc}_e;\Sigma^{\outc}_e;\omega_e)$
\\
Our goal is to show that $\Delta;\Gamma;\Phi^{\inc};\Sigma^{\inc} \vdash e_f[x \mapsto e] : \omega \dashv \Phi^{\outc};\Sigma^{\outc}$. The proof proceeds by induction on the typing derivation of $\Delta;\Gamma;\Phi^{\inc},\Phi^{\inc}_x;\Sigma^{\inc},x:\omega^{\inc}_x \vdash e_f : \omega_f \dashv \Phi_f^{\outc};\Sigma_f^{\outc}, x:\omega^{\outc}_x$.
\begin{itemize}
  \item \textbf{Case} $e_f = x$ \\
  This case is trivial by $\ruleName{T-S-Var}$.
  \item \textbf{Case} $e_f = e_g \scApp e_y$
  \begin{itemize}
    \item \textbf{When} $x$ is linearly used in $e_y$,
    by inversion on $\ruleName{T-S-App}$, we know that:
    \\
    $\Delta;\Gamma;\Phi^{\inc},\Phi^{\inc}_x;\Sigma^{\inc},x:\omega^{\inc}_x \vdash e_g: \omega_g \scFun \omega_h \dashv \Phi^{\outc}_g;\Sigma^{\outc}_g$
    \\
    $\Delta;\Gamma;\Phi^{\inc},\Phi^{\inc}_x;\Sigma^{\inc},x:\omega^{\inc}_x \vdash e_y: \omega_y \dashv \Phi^{\outc}_y;\Sigma^{\outc}_y, x:\omega^{y}_x$
    \\
    $\mathbf{erase}(\omega_g) = \mathbf{erase}(\omega_y)$
    \\
    $(\Phi^{\outc}_f;\Sigma^{\outc}_f, x:\omega^{\outc}_x;\omega_f) = \mathbf{compTrace}(\Phi^{\outc}_g;\Sigma^{\outc}_g;\omega_g;\omega_h;\Phi^{\outc}_y;\Sigma^{\outc}_y, x:\omega^{y}_x;\omega_y)$
    \\
    By the induction hypothesis we know that:
    \\
    $(\Phi^{\outc}_u;\Sigma^{\outc}_u;\omega^{\outc}_y) = \mathbf{compTrace}(\Phi^{\outc}_y;\Sigma^{\outc}_y;\omega^{y}_x;\omega_y;\Phi^{\outc}_e;\Sigma^{\outc}_e;\omega_e)$
    \\
    $\Delta;\Gamma;\Phi^{\inc};\Sigma^{\inc} \vdash e_y[x \mapsto e]: \omega^{\outc}_y \dashv \Phi^{\outc}_u;\Sigma^{\outc}_u$
    \\
    Our goal is to show that $(\Phi^{\outc};\Sigma^{\outc};\omega) = \mathbf{compTrace}(\Phi^{\outc}_g;\Sigma^{\outc}_g;\omega_g;\omega_h;\Phi^{\outc}_u;\Sigma^{\outc}_u;\omega^{\outc}_y)$.
    \\
    Since the operation in $\mathbf{compTrace}$ is uniform for each trace during the enumeration, it is sufficient to consider a single trace (up to renaming) for the proof. We use a tilde on the name ($\tilde{\omega}_e$, for example) to indicate that only a single trace is considered. 
    \begin{itemize}
      \item \textbf{For} each trace which is present in $\Phi^{\outc}$, listed below is the evidence we currently know:
      \\
      $\tilde{\omega}_e[\overline{\mathcal{S}}_e] := \tilde{\omega}^{\outc}_x[\overline{\mathcal{S}}_e] \quad \tilde{\omega}_f[\overline{\mathcal{S}}_e] = \tilde{\omega} \quad (\tilde{\Sigma}^{\outc}_f,\tilde{\Sigma}^{\outc}_e)[\overline{\mathcal{S}}_e] = \tilde{\Sigma}^{\outc}$
      \\
      $\tilde{\omega}_g[\overline{\mathcal{S}}_g] := \tilde{\omega}_y[\overline{\mathcal{S}}_g] \quad \tilde{\omega}_h[\overline{\mathcal{S}}_g] = \tilde{\omega}_f \quad \tilde{\omega}^{y}_x[\overline{\mathcal{S}}_g] = \tilde{\omega}^{\outc}_x \quad (\tilde{\Sigma}^{\outc}_g,\tilde{\Sigma}^{\outc}_y)[\overline{\mathcal{S}}_g] = \tilde{\Sigma}^{\outc}_f$
      \\
      $\tilde{\omega}_e[\overline{\mathcal{S}}_y] := \tilde{\omega}^{y}_x[\overline{\mathcal{S}}_y] \quad \tilde{\omega}_y[\overline{\mathcal{S}}_y] = \tilde{\omega}^{\outc}_y \quad (\tilde{\Sigma}^{\outc}_y,\tilde{\Sigma}^{\outc}_e)[\overline{\mathcal{S}}_y] = \tilde{\Sigma}^{\outc}_u$
      \\
      We write $:=$ in some equations to indicate that the equation also defines a substitution. For example, $\tilde{\omega}_e[\overline{\mathcal{S}}_e] := \tilde{\omega}^{\outc}_x[\overline{\mathcal{S}}_e]$ means $\overline{\mathcal{S}}_e$ is defined by the trace unification of $\tilde{\omega}_e$ and $\tilde{\omega}^{\outc}_x$ using $\mathbf{unifyTrace}$. We need to show that $\mathbf{unifyTrace}$ gives us a substitution $\overline{\mathcal{S}}_h$ (if the unification is successful), such that $\tilde{\omega}_g[\overline{\mathcal{S}}_h] := \tilde{\omega}^{\outc}_y[\overline{\mathcal{S}}_h]$, $\tilde{\omega}_h[\overline{\mathcal{S}}_h] = \tilde{\omega}$, and $(\tilde{\Sigma}^{\outc}_g,\tilde{\Sigma}^{\outc}_u)[\overline{\mathcal{S}}_h] = \tilde{\Sigma}^{\outc}$, that is: \\
      $\tilde{\omega}_g[\overline{\mathcal{S}}_h] := \tilde{\omega}_y[\overline{\mathcal{S}}_y][\overline{\mathcal{S}}_h] \quad \tilde{\omega}_h[\overline{\mathcal{S}}_h] = \tilde{\omega}_h[\overline{\mathcal{S}}_g][\overline{\mathcal{S}}_e]$
      \\
      $(\tilde{\Sigma}^{\outc}_g,(\tilde{\Sigma}^{\outc}_y,\tilde{\Sigma}^{\outc}_e)[\overline{\mathcal{S}}_y])[\overline{\mathcal{S}}_h] = ((\tilde{\Sigma}^{\outc}_g,\tilde{\Sigma}^{\outc}_y)[\overline{\mathcal{S}}_g],\tilde{\Sigma}^{\outc}_e)[\overline{\mathcal{S}}_e]$

      \begin{itemize}
        \item \textbf{For} showing that $\tilde{\omega}_h[\overline{\mathcal{S}}_h] = \tilde{\omega}_h[\overline{\mathcal{S}}_g][\overline{\mathcal{S}}_e]$ with $\overline{\mathcal{S}}_h$ defined as $\tilde{\omega}_g[\overline{\mathcal{S}}_h] := \tilde{\omega}_y[\overline{\mathcal{S}}_y][\overline{\mathcal{S}}_h]$, we need to look into the definition of $\overline{\mathcal{S}}_h$ and $\overline{\mathcal{S}}_e$.
        \begin{itemize}
          \item \textbf{For} the definition of $\overline{\mathcal{S}}_h$, we know that:
          \\
          $dom(\overline{\mathcal{S}}_y) \subseteq mems(\tilde{\omega}_e) \cup mems(\tilde{\omega}^{y}_x) \quad mems(\tilde{\omega}_e) \cap mems(\tilde{\omega}^{y}_x) = \emptyset$
          \\
          $dom(\overline{\mathcal{S}}_g) \subseteq mems(\tilde{\omega}_g) \cup mems(\tilde{\omega}_y) \quad mems(\tilde{\omega}_g) \cap mems(\tilde{\omega}_y) = \emptyset$
          \\
          $mems(\tilde{\omega}_y) \cap mems(\tilde{\omega}_e) = \emptyset \quad mems(\tilde{\omega}_g) \cap mems(\tilde{\omega}_e) = \emptyset$
          \\
          $mems(\tilde{\omega}_y) \cap mems(\tilde{\omega}^{y}_x) \supseteq \emptyset$
          \\
          Thus, $mems(\overline{\mathcal{S}}_y(dom(\overline{\mathcal{S}}_y) \cap dom(\overline{\mathcal{S}}_g))) \cap (dom(\overline{\mathcal{S}}_g) \cup mems(\tilde{\omega}_y)) = \emptyset$,
          \\
          and by the definition of $\mathbf{unifyTrace}$ we can split $\overline{\mathcal{S}}_h$ to two stages, named as $\overline{\mathcal{S}}^{yg}_h$ and $\overline{\mathcal{S}}^{u}_h$, where $\overline{\mathcal{S}}^{yg}_h$ can be defined as follows based on $\overline{\mathcal{S}}_g$.
          \\
          $\overline{\mathcal{S}}^{yg}_h = \forall \tau_{\bunch{\beta}} \in dom(\overline{\mathcal{S}}_g) \setminus dom(\overline{\mathcal{S}}_y), \tau_{\bunch{\beta}} \mapsto \overline{\mathcal{S}}_y(\overline{\mathcal{S}}_g(\tau_{\bunch{\beta}}))$.
          \\
          As for $\overline{\mathcal{S}}^{u}_h$, we can deduce that
          \\
          $dom(\overline{\mathcal{S}}^{u}_h) \subseteq mems(\overline{\mathcal{S}}_y(dom(\overline{\mathcal{S}}_g) \cap dom(\overline{\mathcal{S}}_y))) \cup mems(\overline{\mathcal{S}}_g(dom(\overline{\mathcal{S}}_g) \cap dom(\overline{\mathcal{S}}_y)))$,
          \\
          and for all $\tau_{\bunch{\beta}} \in dom(\overline{\mathcal{S}}^{u}_h)$, $\overline{\mathcal{S}}_y(\tau_{\bunch{\beta}})[\overline{\mathcal{S}}^{u}_h] = \overline{\mathcal{S}}_g(\tau_{\bunch{\beta}})[\overline{\mathcal{S}}^{u}_h]$.
          \\
          Finally, we get $\tilde{\omega}_g[\overline{\mathcal{S}}^{yg}_h][\overline{\mathcal{S}}^{u}_h] = \tilde{\omega}_y[\overline{\mathcal{S}}_y][\overline{\mathcal{S}}^{yg}_h][\overline{\mathcal{S}}^{u}_h]$.
    
          \item \textbf{For} the definition of $\overline{\mathcal{S}}_e$, from the evidence we know that $\tilde{\omega}_e[\overline{\mathcal{S}}_e] := \tilde{\omega}^{y}_x[\overline{\mathcal{S}}_g][\overline{\mathcal{S}}_e]$.
          Similarly, we have\\ $mems(\overline{\mathcal{S}}_g(dom(\overline{\mathcal{S}}_g) \cap dom(\overline{\mathcal{S}}_y))) \cap (dom(\overline{\mathcal{S}}_y) \cup mems(\tilde{\omega}^{y}_x)) = \emptyset$, and by the definition of $\mathbf{unifyTrace}$ we can split $\overline{\mathcal{S}}_e$ to two stages, named as $\overline{\mathcal{S}}^{gy}_e$ and $\overline{\mathcal{S}}^{u}_e$, where $\overline{\mathcal{S}}^{gy}_e$ can be defined as follows based on $\overline{\mathcal{S}}_y$.
          \\
          $\overline{\mathcal{S}}^{gy}_e = \forall \tau_{\bunch{\beta}} \in dom(\overline{\mathcal{S}}_y) \setminus dom(\overline{\mathcal{S}}_g), \tau_{\bunch{\beta}} \mapsto \overline{\mathcal{S}}_g(\overline{\mathcal{S}}_y(\tau_{\bunch{\beta}}))$.
          \\
          As for $\overline{\mathcal{S}}^{u}_e$, we can deduce that
          \\
          $dom(\overline{\mathcal{S}}^{u}_e) \subseteq mems(\overline{\mathcal{S}}_y(dom(\overline{\mathcal{S}}_g) \cap dom(\overline{\mathcal{S}}_y))) \cup mems(\overline{\mathcal{S}}_g(dom(\overline{\mathcal{S}}_g) \cap dom(\overline{\mathcal{S}}_y)))$
          \\
          and for all $\tau_{\bunch{\beta}} \in dom(\overline{\mathcal{S}}^{u}_e)$, $\overline{\mathcal{S}}_y(\tau_{\bunch{\beta}})[\overline{\mathcal{S}}^{u}_e] = \overline{\mathcal{S}}_g(\tau_{\bunch{\beta}})[\overline{\mathcal{S}}^{u}_e]$.
          \\
          Finally, we get $\tilde{\omega}_e[\overline{\mathcal{S}}^{gy}_e][\overline{\mathcal{S}}^{u}_e] = \tilde{\omega}^{y}_x[\overline{\mathcal{S}}_g][\overline{\mathcal{S}}^{gy}_e][\overline{\mathcal{S}}^{u}_e]$.
        \end{itemize}
        Now we need to show that $\tilde{\omega}_h[\overline{\mathcal{S}}^{yg}_h][\overline{\mathcal{S}}^{u}_h] = \tilde{\omega}_h[\overline{\mathcal{S}}_g][\overline{\mathcal{S}}^{gy}_e][\overline{\mathcal{S}}^{u}_e]$.\\
        Since $mems(\tilde{\omega}_h) \cap dom(\overline{\mathcal{S}}_y) = \emptyset$, $\overline{\mathcal{S}}^{yg}_h$ can be broken down, and we get $\tilde{\omega}_h[\overline{\mathcal{S}}_g][\overline{\mathcal{S}}_y][\overline{\mathcal{S}}^{u}_h]$. On the other hand, since $mems(\tilde{\omega}_h[\overline{\mathcal{S}}_g]) \cap dom(\overline{\mathcal{S}}_g) = \emptyset$ and $mems(\tilde{\omega}_h[\overline{\mathcal{S}}_g][\overline{\mathcal{S}}_y]) \cap dom(\overline{\mathcal{S}}_g) = \emptyset$, $\overline{\mathcal{S}}^{gy}_e$ can be replaced by $\overline{\mathcal{S}}_y$, and we get $\tilde{\omega}_h[\overline{\mathcal{S}}_g][\overline{\mathcal{S}}_y][\overline{\mathcal{S}}^{u}_e]$. Finally, $\overline{\mathcal{S}}^{u}_h$ and $\overline{\mathcal{S}}^{u}_e$ are identical by their definitions, so we get the equation $\tilde{\omega}_h[\overline{\mathcal{S}}_g][\overline{\mathcal{S}}_y][\overline{\mathcal{S}}^{u}_h] = \tilde{\omega}_h[\overline{\mathcal{S}}_g][\overline{\mathcal{S}}_y][\overline{\mathcal{S}}^{u}_h]$, that is, $\tilde{\omega}_h[\overline{\mathcal{S}}_h] = \tilde{\omega}_h[\overline{\mathcal{S}}_g][\overline{\mathcal{S}}_e]$.

        \item \textbf{For} showing that $(\tilde{\Sigma}^{\outc}_g,(\tilde{\Sigma}^{\outc}_y,\tilde{\Sigma}^{\outc}_e)[\overline{\mathcal{S}}_y])[\overline{\mathcal{S}}_h] = ((\tilde{\Sigma}^{\outc}_g,\tilde{\Sigma}^{\outc}_y)[\overline{\mathcal{S}}_g],\tilde{\Sigma}^{\outc}_e)[\overline{\mathcal{S}}_e]$, we can decompose the equation into three parts as follows.
        \begin{itemize}
          \item $\tilde{\Sigma}^{\outc}_g[\overline{\mathcal{S}}_h] = \tilde{\Sigma}^{\outc}_g[\overline{\mathcal{S}}_g][\overline{\mathcal{S}}_e]$ can be analyzed in a similar way to $\tilde{\omega}_h[\overline{\mathcal{S}}_h] = \tilde{\omega}_h[\overline{\mathcal{S}}_g][\overline{\mathcal{S}}_e]$.

          \item $\tilde{\Sigma}^{\outc}_y[\overline{\mathcal{S}}_y][\overline{\mathcal{S}}_h] = \tilde{\Sigma}^{\outc}_y[\overline{\mathcal{S}}_g][\overline{\mathcal{S}}_e]$ can be written as\\ 
          $\tilde{\Sigma}^{\outc}_y[\overline{\mathcal{S}}_y][\overline{\mathcal{S}}^{yg}_h][\overline{\mathcal{S}}^{u}_h] = \tilde{\Sigma}^{\outc}_y[\overline{\mathcal{S}}_g][\overline{\mathcal{S}}^{gy}_e][\overline{\mathcal{S}}^{u}_e]$.
          By representing $\overline{\mathcal{S}}_y$ as $\overline{\mathcal{S}}^g_y \circ \overline{\mathcal{S}}^y_y$, where $\overline{\mathcal{S}}^g_y = \forall \tau_{\bunch{\beta}} \in dom(\overline{\mathcal{S}}_y) \cap dom(\overline{\mathcal{S}}_g), \tau_{\bunch{\beta}} \mapsto \overline{\mathcal{S}}_y(\tau_{\bunch{\beta}})$ and $\overline{\mathcal{S}}^y_y = \forall \tau_{\bunch{\beta}} \in dom(\overline{\mathcal{S}}_y) \setminus dom(\overline{\mathcal{S}}_g), \tau_{\bunch{\beta}} \mapsto \overline{\mathcal{S}}_y(\tau_{\bunch{\beta}})$ and similarly representing $\overline{\mathcal{S}}_g$ as $\overline{\mathcal{S}}^y_g \circ \overline{\mathcal{S}}^g_g$, we get $\tilde{\Sigma}^{\outc}_y[\overline{\mathcal{S}}^g_y][\overline{\mathcal{S}}^y_y][\overline{\mathcal{S}}^{yg}_h][\overline{\mathcal{S}}^{u}_h] = \tilde{\Sigma}^{\outc}_y[\overline{\mathcal{S}}^y_g][\overline{\mathcal{S}}^g_g][\overline{\mathcal{S}}^{gy}_e][\overline{\mathcal{S}}^{u}_e]$.
          \\
          For $\overline{\mathcal{S}}^{yg}_h$, $dom(\overline{\mathcal{S}}^{yg}_h) \cap mems(\tilde{\Sigma}^{\outc}_y) = dom(\overline{\mathcal{S}}^{yg}_h) \cap mems(\tilde{\Sigma}^{\outc}_y[\overline{\mathcal{S}}_y])$. Besides, for all $\tau_{\bunch{\beta}} \in (dom(\overline{\mathcal{S}}_g) \setminus dom(\overline{\mathcal{S}}_y)) \cap mems(\tilde{\Sigma}^{\outc}_y)$, there is $\overline{\mathcal{S}}_y(\overline{\mathcal{S}}_g(\tau_{\bunch{\beta}})) = \overline{\mathcal{S}}_g(\tau_{\bunch{\beta}})$ and $mems(\overline{\mathcal{S}}_g(\tau_{\bunch{\beta}})) \cap dom(\overline{\mathcal{S}}_y) = \emptyset$. Thus, $\overline{\mathcal{S}}^{yg}_h$ here can be replaced by $\overline{\mathcal{S}}^g_g$ and its position can be swapped with $\overline{\mathcal{S}}^g_y$ and $\overline{\mathcal{S}}^y_y$. We can do similarly for $\overline{\mathcal{S}}^{gy}_e$ and replace it with $\overline{\mathcal{S}}^y_y$.\\
          Then we get $\tilde{\Sigma}^{\outc}_y[\overline{\mathcal{S}}^y_y][\overline{\mathcal{S}}^g_g][\overline{\mathcal{S}}^g_y][\overline{\mathcal{S}}^{u}_h] = \tilde{\Sigma}^{\outc}_y[\overline{\mathcal{S}}^y_y][\overline{\mathcal{S}}^g_g][\overline{\mathcal{S}}^y_g][\overline{\mathcal{S}}^{u}_e]$. By the definition of $\overline{\mathcal{S}}^{u}_h$ and $\overline{\mathcal{S}}^{u}_e$, this equation holds.

          \item $\tilde{\Sigma}^{\outc}_e[\overline{\mathcal{S}}_y][\overline{\mathcal{S}}_h] = \tilde{\Sigma}^{\outc}_e[\overline{\mathcal{S}}_e]$ can be written as  $\tilde{\Sigma}^{\outc}_e[\overline{\mathcal{S}}_y][\overline{\mathcal{S}}^{yg}_h][\overline{\mathcal{S}}^{u}_h] = \tilde{\Sigma}^{\outc}_e[\overline{\mathcal{S}}^{gy}_e][\overline{\mathcal{S}}^{u}_e]$, and it can be analyzed in a similar way to $\tilde{\omega}_h[\overline{\mathcal{S}}_h] = \tilde{\omega}_h[\overline{\mathcal{S}}_g][\overline{\mathcal{S}}_e]$.
        \end{itemize}
      \end{itemize}

      \item \textbf{For} each trace which is not present in $\Phi^{\outc}$ because in the corresponding $\mathbf{compTrace}$ iteration the $\mathbf{unifyTrace}$ function returns $\mathbf{failTrace}$
      \begin{itemize}
        \item \textbf{If} $\overline{\mathcal{S}}_g$ succeeds but $\overline{\mathcal{S}}_e$ fails, $\overline{\mathcal{S}}_y$ can
        \begin{itemize}
          \item \textbf{Fail}. In this case, the corresponding trace won't exists.
          \item \textbf{Succeed}. In this case, $\overline{\mathcal{S}}_h$ must fail. If both $\overline{\mathcal{S}}_g$ and $\overline{\mathcal{S}}_y$ succeed, then it is possible to define $\overline{\mathcal{S}}^{gy}_e$ as part of $\overline{\mathcal{S}}_e$, and the failure of $\overline{\mathcal{S}}^u_e$ becomes the only reason for $\overline{\mathcal{S}}_e$ to fail. If $\overline{\mathcal{S}}^u_e$ fails, $\overline{\mathcal{S}}^u_h$ which share the same definition with $\overline{\mathcal{S}}^u_e$ will also fail. Thus, $\overline{\mathcal{S}}_h$ which depends on $\overline{\mathcal{S}}^u_h$ must fail.
        \end{itemize}
        \item \textbf{If} $\overline{\mathcal{S}}_g$ fails, then $\overline{\mathcal{S}}_h$ must fail because if $\overline{\mathcal{S}}_g$ which unifies $\tilde{\omega}_g$ and $\tilde{\omega}_y$ fails, it is impossible for $\overline{\mathcal{S}}_h$ which unifies $\tilde{\omega}_g$ and $\tilde{\omega}_y[\overline{\mathcal{S}}_h]$ to succeed.
      \end{itemize}
    \end{itemize}
    \item \textbf{When} $x$ is linearly used in $e_g$, the proof is similar.
  \end{itemize}
  \item \textbf{Case} $e_f = \keyword{st} \ y \scFun e_b$ \\
  We need to show that $\Delta;\Gamma;\Phi^{\inc};\Sigma^{\inc} \vdash (\keyword{st} \ y \scFun e_b)[x \mapsto e] : \omega \dashv \Phi^{\outc};\Sigma^{\outc}$. By linearity, we know that $y \neq x$, $x$ is linearly used in $e_b$, and so is $y$. Thus, by inversion on $\ruleName{T-S-Lam}$, we know that:\\
  $\Delta;\Gamma;\Phi^{\inc},\Phi^{\inc}_x, \Phi^{\inc}_y;\Sigma^{\inc},x:\omega^{\inc}_x, y:\omega^{\inc}_y \vdash e_b : \omega_b \dashv \Phi_f^{\outc};\Sigma_f^{\outc}, x:\omega^{\outc}_x, y:\omega^{\outc}_y$
  \\
  $(\Phi^{\outc};\Sigma^{\outc};\omega) = \mathbf{compTrace}(\Phi^{\outc}_f;\Sigma^{\outc}_f;\omega^{\outc}_x;\omega^{\outc}_y \scFun \omega_b;\Phi^{\outc}_e;\Sigma^{\outc}_e;\omega_e)$
  \\
  and by the induction hypothesis we know that:\\
  $(\Phi^{\outc}_b;\Sigma^{\outc}_b, y:\omega^{b}_y;\omega^{\outc}_b) = \mathbf{compTrace}(\Phi^{\outc}_f;\Sigma^{\outc}_f, y:\omega^{\outc}_y;\omega^{\outc}_x;\omega_b;\Phi^{\outc}_e;\Sigma^{\outc}_e;\omega_e)$
  \\
  $\Delta;\Gamma;\Phi^{\inc}, \Phi^{\inc}_y;\Sigma^{\inc}, y:\omega^{\inc}_y \vdash e_b[x \mapsto e]: \omega^{\outc}_b \dashv \Phi^{\outc}_b;\Sigma^{\outc}_b, y:\omega^{b}_y$
  \\
  By the definition of $\mathbf{compTrace}$, we know that $\Phi^{\outc}_b = \Phi^{\outc}$, $\Sigma^{\outc}_b = \Sigma^{\outc}$, and $(\omega^{b}_y \scFun \omega^{\outc}_b) = \omega$.
  \\
  Then by $\ruleName{T-S-Lam}$, we know that $\Delta;\Gamma;\Phi^{\inc};\Sigma^{\inc} \vdash \keyword{st} \ y \scFun e_b[x \mapsto e]: \omega^{b}_y \scFun \omega^{\outc}_b \dashv \Phi^{\outc}_b;\Sigma^{\outc}_b$, which gives us the proof goal.
\end{itemize}
\end{proofEnd}
\end{proof}

\begin{theoremEnd}[all end]{lemma}[Seq Reduction]
\label{lem:seq}
If $\Delta;\Gamma \vdash (\keyword{seq} \scApp e_m \scApp e_j) \gets e_i : \sRes{\varphi_r}{\tau_r} \dashv \Phi_r$, then $\Delta;\Gamma \vdash e_j \gets (e_m \gets e_i) : \sRes{\varphi_r}{\tau_r} \dashv \Phi_r$.
\end{theoremEnd}
\begin{proofEnd}
By inversion on $\ruleName{T-Seq}$, $\ruleName{T-S-App}$ and $\ruleName{T-Exec}$, we know:
\\
$\Delta;\Gamma \vdash e_m: \sFun{\varphi_m}{\tau_m}{\tau_n} \dashv \Phi_m \quad \Delta;\Gamma \vdash e_j: \sFun{\varphi_j}{\tau_j}{\tau_k} \dashv \Phi_j \quad \Delta;\Gamma \vdash e_i: \sRes{\varphi_i}{\tau_i} \dashv \varphi_i \cstOf \bunch{}$
\\
$\mathbf{erase}(\tau_n) = \mathbf{erase}(\tau_j) \quad \mathbf{erase}(\tau_m) = \mathbf{erase}(\tau_i) \quad \mathbf{erase}(\tau_k) = \mathbf{erase}(\tau_r)$
\\
The full expansion of the $\mathbf{compTrace}$ computation is too complex to show. Alternatively, since the operation in $\mathbf{compTrace}$ is uniform for each trace during the enumeration, it is sufficient to consider a single trace (up to renaming) for the proof. We use a tilde on the name ($\tilde{\tau}_m$, for example) to indicate that only a single trace is considered.
\begin{itemize}
  \item \textbf{For} each trace which is present in $\Phi_r$, we have \\
  $\tilde{\tau}_n[\overline{\mathcal{S}}_{jn}] := \tilde{\tau}_j[\overline{\mathcal{S}}_{jn}] \quad \tilde{\tau}_m[\overline{\mathcal{S}}_{jn}][\overline{\mathcal{S}}_{imjn}] := \tilde{\tau}_i \quad \tilde{\tau}_k[\overline{\mathcal{S}}_{jn}][\overline{\mathcal{S}}_{imjn}] = \tilde{\tau}_r$
  \\
  We write $:=$ in some equations to indicate that the equation also defines a substitution. For example, $\tilde{\tau}_n[\overline{\mathcal{S}}_{jn}] := \tilde{\tau}_j[\overline{\mathcal{S}}_{jn}]$ means $\overline{\mathcal{S}}_{jn}$ is defined by the trace unification of $\tilde{\tau}_n$ and $\tilde{\tau}_j$ using $\mathbf{unifyTrace}$. According the evidence we know, there must be a $\overline{\mathcal{S}}_{im}$ such that $\tilde{\tau}_m[\overline{\mathcal{S}}_{im}] := \tilde{\tau}_i$ because if we have $\tilde{\tau}_m[\overline{\mathcal{S}}_{jn}][\overline{\mathcal{S}}_{imjn}] := \tilde{\tau}_i$, by the definition of $\mathbf{unifyTrace}$, we must be able to get such $\overline{\mathcal{S}}_{im}$. Furthermore, there must also be a $\overline{\mathcal{S}}_{nimj}$ such that $\tilde{\tau}_j[\overline{\mathcal{S}}_{nimj}] := \tilde{\tau}_n[\overline{\mathcal{S}}_{im}]$ because we already know that $\tilde{\tau}_m[\overline{\mathcal{S}}_{jn}][\overline{\mathcal{S}}_{imjn}] = \tilde{\tau}_i = \tilde{\tau}_m[\overline{\mathcal{S}}_{im}]$, and we can get $\tilde{\tau}_n[\overline{\mathcal{S}}_{jn}][\overline{\mathcal{S}}_{imjn}] = \tilde{\tau}_n[\overline{\mathcal{S}}_{im}]$ because $mems(\tilde{\tau}_n) \subseteq mems(\tilde{\tau}_m)$, then by applying $\overline{\mathcal{S}}_{imjn}$ on both sides of $\tilde{\tau}_n[\overline{\mathcal{S}}_{jn}] = \tilde{\tau}_j[\overline{\mathcal{S}}_{jn}]$, we get $\tilde{\tau}_n[\overline{\mathcal{S}}_{jn}][\overline{\mathcal{S}}_{imjn}] = \tilde{\tau}_j[\overline{\mathcal{S}}_{jn}][\overline{\mathcal{S}}_{imjn}]$, and we get $\tilde{\tau}_j[\overline{\mathcal{S}}_{jn}][\overline{\mathcal{S}}_{imjn}] = \tilde{\tau}_j[\overline{\mathcal{S}}_{nimj}]$ by transitivity, and finally we get $\tilde{\tau}_k[\overline{\mathcal{S}}_{jn}][\overline{\mathcal{S}}_{imjn}] = \tilde{\tau}_k[\overline{\mathcal{S}}_{nimj}]$ because $mems(\tilde{\tau}_k) \subseteq mems(\tilde{\tau}_j)$. This also gives us the proof goal, that is $\tilde{\tau}_k[\overline{\mathcal{S}}_{nimj}] = \tilde{\tau}_r$.
  \item \textbf{For} each trace which is not present in $\Phi^{\outc}$ because in the corresponding $\mathbf{compTrace}$ iteration the $\mathbf{unifyTrace}$ function returns $\mathbf{failTrace}$. 
  \begin{itemize}
    \item \textbf{If} $\overline{\mathcal{S}}_{jn}$ succeeds and $\overline{\mathcal{S}}_{imjn}$ fails, then $\overline{\mathcal{S}}_{im}$ can
    \begin{itemize}
      \item \textbf{Fail}. In this case, the corresponding trace won't exists.
      \item \textbf{Succeed}. This case is only possible when $\overline{\mathcal{S}}_{imjn}$ fails because for each $\tau_{\bunch{\beta}} \in mems(\tilde{\tau}_m)$, $\overline{\mathcal{S}}_{jn}(\tau_{\bunch{\beta}})$ cannot be unified with $\overline{\mathcal{S}}_{im}(\tau_{\bunch{\beta}})$. In this case, $\overline{\mathcal{S}}_{nimj}$ must also fail because $mems(\tilde{\tau}_n) \subseteq mems(\tilde{\tau}_m)$.
    \end{itemize}

    \item \textbf{If} $\overline{\mathcal{S}}_{jn}$ fails, then $\overline{\mathcal{S}}_{im}$ can
    \begin{itemize}
      \item \textbf{Fail}. In this case, the corresponding trace won't exists.
      \item \textbf{Succeed}. In this case, $\overline{\mathcal{S}}_{nimj}$ must fail because if $\overline{\mathcal{S}}_{jn}$ which unifies $\tilde{\tau}_n$ and $\tilde{\tau}_j$ fails, it is impossible for $\overline{\mathcal{S}}_{nimj}$ which unifies $\tilde{\tau}_n[\overline{\mathcal{S}}_{im}]$ and $\tilde{\tau}_j$ to succeed.
    \end{itemize}
  \end{itemize} 
\end{itemize}
\end{proofEnd}

With the lemmata shown here and with more detail in the appendix, the proofs for the two parts of type soundness, preservation (\Cref{crl:pres}) and progress (\Cref{thm:prog}), are straightforward.

\begin{theorem}[Subject Reduction]
\label{thm:sr}
If $\Delta \vdash e_m:\omega \dashv \Phi$, and $e_m \stepRel e_n$, then $\Delta \vdash e_n:\omega \dashv \Phi$.
\end{theorem}
\begin{proof}
By case analysis on the reduction $e_m \stepRel e_n$. Proof for cases that cannot be routinely proven are provided by \Cref{lem:rlSubst}, \Cref{lem:stSubst} and \Cref{lem:seq}.
\end{proof}

\begin{corollary}[Preservation]
\label{crl:pres}
If $\Delta \vdash e_m:\omega \dashv \Phi$, and $e_m \evalRel e_n$, then $\Delta \vdash e_n:\omega \dashv \Phi$.
\end{corollary}
\begin{proof}
This is a direct corollary of \Cref{thm:sr} by the definition of $\evalRel$.
\end{proof}

\begin{theorem}[Progress]
\label{thm:prog}
If $\Delta \vdash e_m:\omega \dashv \Phi$, then either $e_m$ is a value, or there exists an $e_n$ such that $e_m \evalRel e_n$.
\end{theorem}
\begin{proof}
We need to show that if $\Delta \vdash e_m:\omega \dashv \Phi$, then either $e_m$ is a value, or there exists an evaluation context $E$ and two expressions $e_j$ and $e_k$ such that $e_m = E[e_j]$ and $e_j \stepRel e_k$. The proof routinely proceeds by induction on the typing derivation of $\Delta \vdash e_m:\omega \dashv \Phi$.
\end{proof}

\begin{corollary}[Type Soundness]
\label{crl:ts}
If $\Delta \vdash e_m:\omega \dashv \Phi$, then either $e_m$ is a value, or there exists an $e_n$ such that $e_m \evalRel^{+} e_n$ and $\Delta \vdash e_n:\omega \dashv \Phi$.
\end{corollary}
\begin{proof}
This is a direct corollary of \Cref{crl:pres} and \Cref{thm:prog}.
\end{proof}

Besides type soundness, Typed \elevate also has the strong normalization property. The proof technique used in the strong normalization proof of STLC \cite{STLC-SN} can also be applied here.

\begin{theorem}[Strong Normalization]
\label{thm:sn}
If $\Delta \vdash e:\omega \dashv \Phi$, then $e \evalRel^{\ast} v$ where $v$ is a value.
\end{theorem}
\begin{proof}
It is easy to show that $\Delta;\Gamma \vdash (\keyword{rule} \ p \to e) \gets e_i : \sRes{\varphi_r}{\tau_r} \dashv \Phi_r$ always normalize, so there is no need to consider substitution for variables from $\Theta$. Thus, for a term $e$, we need to show that after substituting strong-normalizing terms for variables from $\Gamma$ (by $\mathcal{S}_{\Gamma}$) and $\Sigma$ (by $\mathcal{S}_{\Sigma}$) in $e$, $e[\mathcal{S}_{\Gamma}][\mathcal{S}_{\Sigma}]$ is also strong-normalizing. This proof proceeds by constructing the logical relation on strong normalizing terms indexed by type $\omega$ and then induction on $e$.
\end{proof}

\subsection{Properties of Well-typed Strategies}
\label{sec:st-prop}

This section gives formal statements about the properties of well-typed strategies. \Cref{crl:emptyRes} and \Cref{crl:emptySt} are about well-typed but empty-traced expressions. \Cref{crl:emptyRes} shows that an expression of an empty-traced result type must evaluate to $\keyword{fail}$. \Cref{crl:emptySt} refines the condition in \Cref{crl:emptyRes} and states that if a strategy is empty-traced, then for any well-typed execution of this strategy, the result will be $\keyword{fail}$. The emptiness check for the tracing environment in $\ruleName{T-Let}$ is supported by \Cref{crl:emptySt} because a strategy known to always fail is useless or unproductive.

\begin{corollary}[Empty-traced Failed Result]
\label{crl:emptyRes}
If $\Delta \vdash e : \sRes{\bunch{}}{\tau}$, then $e \evalRel^{\ast} \keyword{fail}$.
\end{corollary}
\begin{proof}
This is a direct corollary of \Cref{thm:sn} and \Cref{crl:ts} because $\keyword{fail}$ is the only value can be typed with $\sRes{\bunch{}}{\tau}$.
\end{proof}

\begin{corollary}[Empty-traced Unproductive Strategy]
\label{crl:emptySt}
If $\Delta \vdash e_f: \sFun{\bunch{}}{\tau_a}{\tau_b}$, then for any $e_i$ that $\Delta \vdash e_f \gets e_i : \sRes{\phi_r}{\tau_r} \dashv \Phi_r$, there is $(e_f \gets e_i) \evalRel^{+} \keyword{fail}$.
\end{corollary}
\begin{proof}
This is a direct corollary of \Cref{crl:emptyRes} and $\ruleName{T-Exec}$.
\end{proof}

\Cref{lem:rlSucc} shows that well-typed and well-traced rule executions must succeed. Due to the compromise on precision in $\ruleName{T-Exec}$, we use a notation in the form of $\Delta;\Gamma \nvdash e : \sRes{\bunch{}}{\tau}$ here to formally states the meaning of "well-traced", that is, $e$ cannot be typed with the empty-traced type $\sRes{\bunch{}}{\tau}$. Specifically in \Cref{lem:rlSucc}, this means the $\mathbf{compTrace}$ in $\ruleName{T-Exec}$ must return a traced triple with at least one trace. In combination with $\Delta;\Gamma \vdash e_r : \sRes{\varphi_r}{\tau_r} \dashv \varphi_r \cstOf \bunch{}$, we get the complete formal statement of "well-typed and well-traced".

\begin{lemmaE}[Successful Rule]
\label{lem:rlSucc}
If for $e_r = (\keyword{rule} \ p \to e) \gets \keyword{succ} \ v$, there is $\Delta;\Gamma \vdash e_r : \sRes{\varphi_r}{\tau_r} \dashv \varphi_r \cstOf \bunch{}$ and $\Delta;\Gamma \nvdash e_r : \sRes{\bunch{}}{\tau}$, then $e_r \evalRel^{+} \keyword{succ} \ v_s$.
\end{lemmaE}
\begin{proof}
The proof proceeds by induction on the derivation of the typing judgement for $p$.
\begin{proofEnd}
By inversion on $\ruleName{T-Exec}$ and $\ruleName{T-R-Lam}$, we know $\Delta \vDash_{\alpha_f} p: \tau_p \dashv \Phi;\Theta$, and $\Delta;\Gamma;\Phi;\Theta \vdash e : \tau_e$, and $\Delta;\Gamma;\alpha_v \cstOf \bunch{};\star \vdash v: \tau_v$, and $\mathbf{erase}(\tau_p) = \mathbf{erase}(\tau_v)$, and there is a $\overline{\mathcal{S}} \neq \mathbf{failTrace}$ generated by $\mathbf{unifyTrace}(\alpha_f \cstOf \phi_f;\sRes{\bunch{\alpha_f}}{\tau_p};\sRes{\bunch{\alpha_f}}{(\tau_v[\alpha_v \mapsto \alpha_f])})$ because $\Delta;\Gamma \nvdash e_r : \sRes{\bunch{}}{\tau}$. By induction on the typing derivation of $\Delta \vDash_{\alpha_f} p: \tau_p \dashv \Phi;\Theta$ and the definition of $\mathbf{unifyTrace}$, we know that there are no mismatched labels in $p$ and $v$. By the definition of $\mathbf{match}$ we know that mismatched labels is not only a source of $\mathbf{failPat}$, but also the only source. Thus, $\mathbf{match}(p;v)$ must succeed, and $e_r \evalRel \keyword{succ} \ v_s$.
\end{proofEnd}
\end{proof}

\begin{theoremEnd}[all end]{lemma}[Enumeration]
\label{lem:enum}
If $\Delta;\Gamma \vdash v_f: \sFun{\varphi_f}{\tau_p}{\tau_e} \dashv \Phi_f$ and $\Delta;\Gamma \vdash \keyword{succ} \ v_i: \sRes{\bunch{\alpha_i}}{\tau_{i}} \dashv \alpha_i \cstOf \bunch{}$ and $\mathbf{erase}(\tau_p) = \mathbf{erase}(\tau_i)$ and $(\varphi_r \cstOf \bunch{};\cdot;\sRes{\varphi_r}{\tau_r}) = \mathbf{compTrace}(\Phi_f;\cdot;\sRes{\varphi_f}{\tau_p};\sRes{\varphi_f}{\tau_e};\alpha_i \cstOf \bunch{};\cdot;\sRes{\bunch{\alpha_i}}{\tau_{i}})$ where $\varphi_r \neq \bunch{}$, then for each $\alpha_r$ in $\varphi_r$, there exists a reduction $(v_f \gets (\keyword{succ} \ v_i)) \evalRel^{+} \keyword{succ} \ v_r$, and for $(\alpha_r \cstOf \bunch{}; \cdot; \sRes{\bunch{\alpha_r}}{\tau'_{r}}) = \mathbf{Select} \ \bunch{\alpha_r} \ \mathbf{in} \ (\varphi_r \cstOf \bunch{}; \cdot; \sRes{\varphi_r}{\tau_r})$, there is $\Delta;\Gamma;\alpha_r \cstOf \bunch{};\star \vdash v_r: \tau'_r$.
\end{theoremEnd}
\begin{proofEnd}
The proof proceeds by induction on $v_f$. There are three cases to consider: $v_f = (\keyword{rule} \ p \to e)$, $v_f = (\keyword{seq} \scApp v_a \scApp v_b)$, and $v_f = (\keyword{choice} \scApp v_a \scApp v_b)$. The $\keyword{rule}$ case is proven by \Cref{lem:rlSucc}. The $\keyword{choice}$ case can be easily proven by applying the induction hypothesis. The $\keyword{seq}$ case can be proven by applying the induction hypothesis and \Cref{lem:seq}.
\end{proofEnd}

Finally, \Cref{crl:succRw} shows that well-typed and well-traced strategy executions must succeed. More precisely, this should be stated as "there must exist at least one successful execution result" because of non-determinism. This corollary relies on type soundness (\Cref{crl:ts}) and \Cref{lem:enum} which is moved to the appendix because of lengthy proposition. An informal description for \Cref{lem:enum} is that it shows each trace corresponds to a possible execution path in a strategy execution, which can be considered as the generalization of \Cref{lem:rlSucc} to strategies.

\begin{corollary}[Successful Rewrite]
\label{crl:succRw}
If $\Delta \vdash e_f \gets e_i : \sRes{\varphi_r}{\tau_r} \dashv \Phi_r$ and $\Delta \nvdash e_f \gets e_i : \sRes{\bunch{}}{\tau}$, then there exists a reduction $(e_f \gets e_i) \evalRel^{+} \keyword{succ} \ v_s$.
\end{corollary}
\begin{proof}
This is a direct corollary of \Cref{lem:enum} and \Cref{crl:ts}.
\end{proof}


\section{Discussion}
\label{sec:discussion}
In this section, we discuss a central design choice we have made regarding the choice combinator.
In Typed \elevate we only support a non-deterministic version of choice where \elevate and Stratego also offer a deterministic version, called left-choice, that first applies the strategy on the left, and only attempts to apply the strategy on the right if the first fails.

Let us understand our design choice by looking at the strategy \lstinline{e8}, that is defined as a choice composition sequentially followed by another rewrite rule:

\begin{lstlisting}[mathescape]
let e8 = (rule m * n -> n * m || rule 1*v -> v) ; rule m + n -> n + m
\end{lstlisting}

The choice composition contains two rules for multiplications, and has the following type:
\begin{lstlisting}[mathescape, frame=none, aboveskip=0.125em, belowskip=0.25em]
(rule m * n -> n * m || rule 1*v -> v) : $\sFun{\bunch{a,b}}{(a_0 * a_1)~|~(1 * b_0)}{(a_1 * a_0)~|~b_0}$
\end{lstlisting}

The single rewrite rule swaps the operands of an addition, and it has the following type:
\begin{lstlisting}[mathescape, frame=none, aboveskip=0.125em, belowskip=0.25em]
rule m + n -> n + m : $\sFun{\bunch{c}}{c_0 + c_1}{c_1 + c_0}$
\end{lstlisting}

Thus, with the rules of our type system their sequential composition has the following type:
\begin{lstlisting}[mathescape]
let e8 = (rule m * n -> n * m || rule 1*v -> v) ; rule m + n -> n + m : $\sFun{\bunch{d}}{1 * (d_0 + d_1)}{1 * (d_1 + d_0)}$
\end{lstlisting}

There is only one trace identified by $d$ in the final strategy type, reflecting that the strategy expects inputs in the form of $1 * (d_0 + d_1)$.
With such inputs the \Cref{crl:succRw} guarantees, that the overall strategy execution must have one successful execution path, which is \lstinline{rule 1 * v -> v} followed by \lstinline{rule m + n -> n + m}.
However, maybe surprisingly, this would not be true when we use the left-biased deterministic choice combinator!
This strategy execution would always return $\keyword{fail}$.
The left-biased choice combinator starts with applying the first strategy to the input, as the pattern \lstinline{m * n} overlaps with \lstinline{1 * v}, the right strategy is inaccessible at runtime and the left strategy will always be chosen.
But, unfortunately, the left strategy does not compose sequentially with the additional rewrite rules, so the overall strategy execution always fails.
%

If we want to make similar formal statements about the properties of well-typed strategies as in \Cref{sec:st-prop}, but with left-biased choice, we will find that \Cref{crl:emptyRes} and \Cref{crl:emptySt} still hold because they only rely on type soundness and strong normalization, which are not affected by switching to left-choice. On the other hand, \Cref{crl:succRw} will not hold anymore because as shown by the example above, with left-choice, the traces lose their correspondence with the execution paths. In other words, we will not be able to prove \Cref{lem:enum} with left-choice.
To obtain a property stating that well-typed and well-traced strategy executions must succeed, we would need to statically remove the traces containing inaccessible patterns from the types, making the design of the tracing mechanism much more challenging. Besides, for strategy combinators containing left-choice, removing traces of inaccessible patterns would mean that the type of strategy arguments can affect the type of existing strategies in the combinator definition. This may bring composability issues and might make the type soundness proof more challenging.

It is worth noting that the statements about well-typed but empty-traced strategies are more resilient to changes in the type system, as they only rely on type soundness and strong normalization.
Changes such as switching to a deterministic version of choice, allowing untraced fallible conditions in rewrite rule definitions and adding primitive types which cannot be precisely traced, can make the statements about the success of well-typed and well-traced strategy executions no longer hold, but as long as the type system is sound and strong normalizing, we can still state that the execution result of well-typed but empty-traced strategies must be $\keyword{fail}$.

\section{Related Work}
\label{sec:related-work}

\paragraph{Type Systems for Rewrite Systems.}
Term-rewriting systems \citep{dershowitz1985computing} have been shown useful in various applications such as program transformation, languages semantics and computations in theorem-proving systems. In strategy languages designed specifically for program transformations such as ELAN by \citet{borovansky1996elan}, Stratego by \citet{visser1998building} and \citet{martin2008stratego}, and TL by \citet{winter2004special}, the users can define strategies using composition operators to control reduction sequences, which is adapted in the design and implementation of \elevate by \citet{bastianhagedorn2020achieving}. Basic type systems and formalization for strategic term rewriting have been presented by \citet{lammel2003typed} and \citet{kaiser2009an} covering generic traversal. \citet{eelco20gradual} shows the type system of Core Gradual Stratego which use gradual type system to combine the flexibility of dynamic typing and the safety provided by statical typing. \citet{koppel:OASIcs.EVCS.2023.16} uses the Compositional Data Types Haskell library, which is initially designed to address the expression problem and used for generic programming \cite{compositional}, to ensure the existence of applicable language constructs during rewriting and achieve type-preserving program transformations. Meanwhile, our contribution focuses on the concrete shape of rewritten programs, emphasizes the productive composition of strategies, and rules out useless strategies statically using a type system with traces.

\paragraph{Row Polymorphic Languages.}
Introduced by \citet{remy1989type} and \citet{wand1991type}, row polymorphism is a parametric polymorphism allows representing an extensible structure in types as a row, usually a sequence of label-type pairs (while other notions exist such as by \citet{morris19row}), which can be used as the basis for (polymorphic) variants and records. Row polymorphism is as expressive as structural subtyping, but works smoothly with Hindley-Milner style type inference, and it has many applications in modeling type systems. The structurally polymorphic types in OCaml are modeled using row polymorphism with guarantees on type soundness \citep{garrigue2015certified}. In Links, \citet{daniel2016liberating} use row polymorphism for the implementation of algebraic effects and effect handlers, providing modular abstraction for effectful computation. \citet{blume06cases} uses rows to realize polymorphic variants and records, and treats cases or pattern matching branches as first-class values so pattern matching expressions can be extended with new cases at any moment. This is similar to composing strategies with the $\keyword{choice}$ combinator in Typed \elevate, but without the precision provided by traces. More recently, \citet{morris19row} introduced a general theory for existing row polymorphic type systems, which focuses on row concatenation and gives rise to the language ROSE as a flexible tool for programming with extensible data types.



\section{Conclusion}
\label{sec:conclusion}

In this paper, we present Typed \elevate, a strategy language with a strong static type system capable of rejecting ill-composed strategies.
We discuss a number of practical strategy compositions, demonstrating the capabilities of our type system to compute precise types for strategy compositions as well as raising warnings and errors for problematic strategy compositions.

Our type system combines a structural type system based on row polymorphism with a novel tracing type system.
We use the structural types to represent the shape of expressions that are rewritten at the type level.
The tracing system is designed as an additional layer on top of the structural types to represent the possible ways a strategy can be executed.

We confirm the soundness of our type system, formally justify why we reject empty-traced strategies (as they are guaranteed to fail at runtime), and show that well-traced strategies must have a possible execution path at runtime.
Finally, we discuss the importance of the non-deterministic choice combinator for our strong formal results.


\bibliography{references}

\newpage
\appendix

\section{Definition of auxiliary functions}

\begin{figure}[H]
  \footnotesize
  \begin{gather*}
  \begin{aligned}
  &\mathbf{p2e}: p \to e \\
  &\mathbf{p2e}(x) = x \\
  &\mathbf{p2e}(\ell \ p) = \ell \ \mathbf{p2e}(p) \\
  &\mathbf{p2e}(()) = () \\
  &\mathbf{p2e}((p_a, p_b)) = (\mathbf{p2e}(p_a), \mathbf{p2e}(p_b))
  \end{aligned}
\end{gather*}
\caption{Definition of pattern-expression conversion}
\label{fig:p2e}
\end{figure}


\section{Definition of trace operations}

\begin{figure}[H]
  \footnotesize
  \begin{gather*}
  \begin{aligned}
  &\mathbf{erase}: t \to t \\
  &\mathbf{erase}(\tau_{\psi}) = \mathbf{erase}(\tau) \\
  &\mathbf{erase}(\nu) = \nu \\
  &\mathbf{erase}(\langle \rho \rangle) = \langle \mathbf{erase}(\rho) \rangle \\
  &\mathbf{erase}(\nu \ \keyword{as} \ \langle \rho \rangle) = \nu \ \keyword{as} \ \langle \mathbf{erase}(\rho) \rangle \\
  &\mathbf{erase}(()) = () \\
  &\mathbf{erase}((\tau_m, \tau_n)) = (\mathbf{erase}(\tau_m), \mathbf{erase}(\tau_n)) \\
  &\mathbf{erase}(\cdot) = \cdot \\
  &\mathbf{erase}(\ell : \tau \mid \rho) = (\ell : \mathbf{erase}(\tau) \mid \mathbf{erase}(\rho)) \\
  &\mathbf{erase}(\sRes{\varphi}{\tau}) = \sRes{\bunch{}}{\mathbf{erase}(\tau)} \\
  &\mathbf{erase}(\sFun{\varphi}{\tau_p}{\tau_e}) = \sFun{\bunch{}}{\mathbf{erase}(\tau_p)}{\mathbf{erase}(\tau_e)} \\
  &\mathbf{erase}((\sFun{\varphi}{\tau_p}{\tau_e}) \scFun \omega) = \mathbf{erase}(\sFun{\varphi}{\tau_p}{\tau_e}) \scFun \mathbf{erase}(\omega)
  \end{aligned}
  \end{gather*}
\caption{Definition of trace erasure}
\label{fig:trace-erasure}
\end{figure}

\begin{figure}[H]
\footnotesize
\begin{gather*}
  \begin{aligned}
    &-[-]: t \to \overline{\mathcal{S}} \to t\\
    &\tau_{\bunch{\beta}}[\overline{\mathcal{S}}] = \overline{\mathcal{S}}(\tau_{\bunch{\beta}}) \ \mathbf{if} \ \tau_{\bunch{\beta}} \in dom(\overline{\mathcal{S}})\\
    &\tau_{\bunch{\beta}}[\overline{\mathcal{S}}] = \tau_{\bunch{\beta}} \ \mathbf{if} \ \tau_{\bunch{\beta}} \notin dom(\overline{\mathcal{S}})\\
    &\nu[\overline{\mathcal{S}}] = \nu \\
    &\langle \rho \rangle[\overline{\mathcal{S}}] = \langle \rho[\overline{\mathcal{S}}] \rangle\\
    &()[\overline{\mathcal{S}}] = () \\
    &(\tau_m, \tau_n)[\overline{\mathcal{S}}] = (\tau_m[\overline{\mathcal{S}}], \tau_n[\overline{\mathcal{S}}]) \\
    &(\ell:\tau \mid \rho)[\overline{\mathcal{S}}] = (\ell:\tau[\overline{\mathcal{S}}] \mid \rho[\overline{\mathcal{S}}]) \\
    &\cdot[\overline{\mathcal{S}}] = \cdot
  \end{aligned}
\end{gather*}
\caption{Definition of trace substitution}
\label{fig:trace-subst}
\end{figure}
  
\begin{figure}[H]
\footnotesize
\begin{gather*}
  \begin{aligned}
    &\mathbf{unifyTrace}: (\Phi;t;t) \to \overline{\mathcal{S}}\\
    &\mathbf{unifyTrace}(\alpha \cstOf \phi;()_{\bunch{\alpha}};()) = \mathbf{failTrace} \\
    &\mathbf{unifyTrace}(\alpha \cstOf \phi;();()_{\bunch{\alpha}}) = \mathbf{failTrace} \\
    &\mathbf{unifyTrace}(\alpha \cstOf \bunch{\beta,\phi};\tau_{\bunch{\beta}};\tau_n) = \mathbf{if} \ \tau_n = \tau \ \mathbf{then} \ \mathbf{failTrace} \ \mathbf{else} \ \tau_{\bunch{\beta}} \mapsto \tau_n \\
    &\mathbf{unifyTrace}(\alpha \cstOf \bunch{\beta,\phi};\tau_m;\tau_{\bunch{\beta}}) = \mathbf{if} \ \tau_m = \tau \ \mathbf{then} \ \mathbf{failTrace} \ \mathbf{else} \ \tau_{\bunch{\beta}} \mapsto \tau_m \\
    &\mathbf{unifyTrace}(\Phi;\langle \rho_m \rangle;\langle \rho_n \rangle) = \mathbf{unifyTrace}(\Phi;\rho_m;\rho_n)\\
    &\mathbf{unifyTrace}(\Phi;(\tau_m, \tau_n); (\tau_j, \tau_k)) = \mathbf{let} \ \overline{\mathcal{S}} = \mathbf{unifyTrace}(\Phi;\tau_m; \tau_j) \\
    &\quad \mathbf{in} \ \mathbf{unifyTrace}(\Phi;\tau_n[\overline{\mathcal{S}}]; \tau_k[\overline{\mathcal{S}}]) \circ \overline{\mathcal{S}}\\
    &\mathbf{unifyTrace}(\Phi;\tau;\tau) = (\cdot \mapsto \cdot) \\
    &\mathbf{unifyTrace}(\Phi;(\ell: \tau_m \mid \rho_m);(\ell: \tau_n \mid \rho_n)) = \mathbf{let} \ \overline{\mathcal{S}} = \mathbf{unifyTrace}(\Phi;\tau_m;\tau_n)\\
    &\quad \mathbf{in} \ \mathbf{unifyTrace}(\Phi;\rho_m[\overline{\mathcal{S}}]; \rho_n[\overline{\mathcal{S}}]) \circ \overline{\mathcal{S}}\\
    &\mathbf{unifyTrace}(\Phi;\rho;\rho) = (\cdot \mapsto \cdot) \\
    &\mathbf{unifyTrace}(\Phi;\sRes{\bunch{\alpha}}{\tau_m};\sRes{\bunch{\alpha}}{\tau_n}) = \mathbf{unifyTrace}(\Phi;\tau_m;\tau_n) \\
    &\mathbf{unifyTrace}(\Phi;\sFun{\bunch{\alpha}}{\tau_m}{\tau_n}; \sFun{\bunch{\alpha}}{\tau_j}{\tau_k}) = \mathbf{let} \ \overline{\mathcal{S}} = \mathbf{unifyTrace}(\Phi;\tau_m; \tau_j) \\
    &\quad \mathbf{in} \ \mathbf{unifyTrace}(\Phi;\tau_n[\overline{\mathcal{S}}]; \tau_k[\overline{\mathcal{S}}]) \circ \overline{\mathcal{S}}
  \end{aligned}
\end{gather*}
\caption{Definition of trace unification}
\label{fig:trace-uni}
\end{figure}

\begin{figure}[H]
  \footnotesize
  \begin{gather*}
  \begin{aligned}
  &\mathbf{traceIds}: (\Phi;\omega) \to \varphi \\
  &\mathbf{traceIds}(\Phi;\sFun{\varphi}{\tau_p}{\tau_e}) = \varphi \ \mathbf{if} \ \Phi \vdash \tau_p \dashv \varphi \ \mathbf{and} \ \Phi \vdash \tau_e \dashv \varphi \\
  &\mathbf{traceIds}(\Phi;\sRes{\varphi}{\tau}) = \varphi \ \mathbf{if} \ \Phi \vdash \tau \dashv \varphi
  \end{aligned}
\end{gather*}
\caption{Definition of $\mathbf{traceIds}$}
\label{fig:traceIds}
\end{figure}

\begin{figure}[H]
  \footnotesize
  \begin{gather*}
  \begin{aligned}
  &\mathbf{Select} \ - \ \mathbf{in} \ -: \varphi \to (\Phi;\Sigma;\omega) \to (\Phi;\Sigma;\omega) \\
  &\mathbf{Select} \ \varphi \ \mathbf{in} \ (\Phi;\Sigma;\omega) = \\
  &\quad (\Phi_r;\Sigma_r;\omega_r) := (\cdot;\mathbf{erase}(\Sigma);\mathbf{erase}(\omega)) \\
  &\quad \mathbf{for} \ \alpha \ \mathbf{in} \ \varphi \\
  &\quad \quad \mathbf{if} \ \alpha \cstOf \phi \in \Phi \\
  &\quad \quad \quad \Sigma_i := \cdot \\
  &\quad \quad \quad \mathbf{for} \ (x:\omega_x) \ \mathbf{in} \ \Sigma \\
  &\quad \quad \quad \quad \Sigma_i := \Sigma_i, x:\mathbf{selectInType}(\Phi;\alpha;\omega_x)\\
  &\quad \quad \quad (\Phi_r;\Sigma_r;\omega_r) := \mathbf{Add} \ (\alpha \cstOf \phi;\Sigma_i;\mathbf{selectInType}(\Phi;\alpha;\omega)) \ \mathbf{to} \ (\Phi_r;\Sigma_r;\omega_r) \\
  &\quad \mathbf{return} \ (\Phi_r;\Sigma_r;\omega_r) \\
  \\
  &\mathbf{selectInType}: (\Phi;\alpha;t) \to t \\
  &\mathbf{selectInType}(\alpha \cstOf \phi, \Phi_r;\alpha;()_{\bunch{\alpha, \psi_r}}) = ()_{\bunch{\alpha}} \\
  &\mathbf{selectInType}(\alpha \cstOf \bunch{\beta, \phi_r}, \Phi_r;\alpha;\tau_{\bunch{\beta, \psi_r}}) = \mathbf{erase}(\tau)_{\bunch{\beta}} \\
  &\mathbf{selectInType}(\alpha \cstOf \phi, \Phi_r;\alpha;\tau_{\psi}) = \mathbf{selectInType}(\alpha \cstOf \phi, \Phi_r;\alpha;\tau) \ \mathbf{if} \ \phi \cap \psi = \emptyset \ \mathbf{and} \ \alpha \notin \psi \\
  &\mathbf{selectInType}(\Phi;\alpha;\langle \rho \rangle) = \langle \mathbf{selectInType}(\Phi;\alpha;\rho) \rangle \\
  &\mathbf{selectInType}(\Phi;\alpha;\nu \ \keyword{as} \ \langle \rho \rangle) = \nu \ \keyword{as} \ \langle \mathbf{selectInType}(\Phi;\alpha;\rho) \rangle\\
  &\mathbf{selectInType}(\Phi;\alpha;\cdot) = \cdot \\
  &\mathbf{selectInType}(\Phi;\alpha;\nu) = \nu \\
  &\mathbf{selectInType}(\Phi;\alpha;(\ell: \tau \mid \rho)) = (\ell : \mathbf{selectInType}(\Phi;\alpha;\tau) \mid \mathbf{selectInType}(\Phi;\alpha;\rho)) \\
  &\mathbf{selectInType}(\Phi;\alpha;()) = () \\
  &\mathbf{selectInType}(\Phi;\alpha;(\tau_m, \tau_n)) = (\mathbf{selectInType}(\Phi;\alpha;\tau_m), \mathbf{selectInType}(\Phi;\alpha;\tau_n)) \\
  &\mathbf{selectInType}(\Phi;\alpha;\sRes{\bunch{\alpha, \varphi_r}}{\tau}) = \sRes{\bunch{\alpha}}{\mathbf{selectInType}(\Phi;\alpha;\tau)} \\
  &\mathbf{selectInType}(\Phi;\alpha;\sRes{\varphi}{\tau}) = \sRes{\bunch{}}{\mathbf{erase}(\tau)} \ \mathbf{if} \ \alpha \notin \varphi \\
  &\mathbf{selectInType}(\Phi;\alpha;\sFun{\bunch{\alpha, \varphi_r}}{\tau_p}{\tau_e}) = \sFun{\bunch{\alpha}}{\mathbf{selectInType}(\Phi;\alpha;\tau_p)}{\mathbf{selectInType}(\Phi;\alpha;\tau_e)} \\
  &\mathbf{selectInType}(\Phi;\alpha;\sFun{\varphi}{\tau_p}{\tau_e}) = \sFun{\bunch{}}{\mathbf{erase}(\tau_p)}{\mathbf{erase}(\tau_e)} \ \mathbf{if} \ \alpha \notin \varphi \\
  &\mathbf{selectInType}(\Phi;\alpha;(\sFun{\varphi}{\tau_p}{\tau_e}) \scFun \omega) = \mathbf{selectInType}(\Phi;\alpha;\sFun{\varphi}{\tau_p}{\tau_e}) \scFun \mathbf{selectInType}(\Phi;\alpha;\omega)
  \end{aligned}
\end{gather*}
\caption{Definition of $\mathbf{Select} \ - \ \mathbf{in} \ -$}
\label{fig:select}
\end{figure}

\begin{figure}[H]
  \footnotesize
  \begin{gather*}
  \begin{aligned}
  &\mathbf{Fresh} \ -: (\Phi;\Sigma;\omega) \to (\Phi;\Sigma;\omega) \\
  &\mathbf{Fresh} \ (\Phi;\Sigma;\omega) = \\
  &\quad (\Phi_r;\Sigma_r;\omega_r) := (\cdot;\mathbf{erase}(\Sigma);\mathbf{erase}(\omega)) \\
  &\quad \mathbf{for} \ \alpha \cstOf \phi \ \mathbf{in} \ \Phi \\
  &\quad \quad (\alpha \cstOf \phi;\Sigma_{\alpha};\omega_{\alpha}) = \mathbf{Select} \ \bunch{\alpha} \ \mathbf{in} \ (\Phi;\Sigma;\omega) \\
  &\quad \quad \phi_i := \mathbf{rmMems}(\alpha \cstOf \phi;\omega_{\alpha};\bunch{}) \\
  &\quad \quad \mathbf{for} \ (x:\omega_x) \ \mathbf{in} \ \Sigma_{\alpha} \\
  &\quad \quad \quad \phi_i := \mathbf{rmMems}(\alpha \cstOf \phi;\omega_x;\phi_i) \\
  &\quad \quad (\Phi_n;\Sigma_n;\omega_n) := (\alpha \cstOf \phi_i;\Sigma_{\alpha};\omega_{\alpha})[\alpha \mapsto \mathbf{fresh} \ \alpha_{n}] \\
  &\quad \quad \mathbf{for} \ \beta \ \mathbf{in} \ \phi_i \\
  &\quad \quad \quad (\Phi_n;\Sigma_n;\omega_n) := (\Phi_n;\Sigma_n;\omega_n)[\beta \mapsto \mathbf{fresh} \ \beta_{n}] \\
  &\quad \quad (\Phi_r;\Sigma_r;\omega_r) := \mathbf{Add} \ (\Phi_n;\Sigma_n;\omega_n) \ \mathbf{to} \ (\Phi_r;\Sigma_r;\omega_r) \\
  &\quad \mathbf{return} \ (\Phi_r;\Sigma_r;\omega_r) \\
  \\
  &\mathbf{rmMems}: (\Phi;t;\phi) \to \phi \\
  &\mathbf{rmMems}(\alpha \cstOf \phi;()_{\bunch{\alpha}};\phi_s) = \phi_s \\
  &\mathbf{rmMems}(\alpha \cstOf \bunch{\beta, \phi_r};\tau_{\bunch{\beta}};\phi_s) = \mathbf{if} \ \beta \in \phi_s \ \mathbf{then} \ \phi_s \ \mathbf{else} \ \bunch{\beta, \phi_s} \\
  &\mathbf{rmMems}(\Phi;\langle \rho \rangle;\phi_s) = \mathbf{rmMems}(\Phi;\rho;\phi_s) \\
  &\mathbf{rmMems}(\Phi;\nu \ \keyword{as} \ \langle \rho \rangle;\phi_s) = \mathbf{rmMems}(\Phi;\rho;\phi_s) \\
  &\mathbf{rmMems}(\Phi;\cdot;\phi_s) = \phi_s \\
  &\mathbf{rmMems}(\Phi;\nu;\phi_s) = \phi_s \\
  &\mathbf{rmMems}(\Phi;(\ell: \tau \mid \rho);\phi_s) = \mathbf{rmMems}(\Phi;\rho;\mathbf{rmMems}(\Phi;\tau;\phi_s)) \\
  &\mathbf{rmMems}(\Phi;();\phi_s) = \phi_s \\
  &\mathbf{rmMems}(\Phi;(\tau_m, \tau_n);\phi_s) = \mathbf{rmMems}(\Phi;\tau_n;\mathbf{rmMems}(\Phi;\tau_m;\phi_s)) \\
  &\mathbf{rmMems}(\Phi;\sRes{\varphi}{\tau};\phi_s) = \mathbf{rmMems}(\Phi;\tau;\phi_s) \\
  &\mathbf{rmMems}(\Phi;\sFun{\varphi}{\tau_p}{\tau_e};\phi_s) = \mathbf{rmMems}(\Phi;\tau_e;\mathbf{rmMems}(\Phi;\tau_p;\phi_s)) \\
  &\mathbf{rmMems}(\Phi;(\sFun{\varphi}{\tau_p}{\tau_e}) \scFun \omega;\phi_s) = \mathbf{rmMems}(\Phi;\omega;\mathbf{rmMems}(\Phi;\sFun{\varphi}{\tau_p}{\tau_e};\phi_s))
  \end{aligned}
\end{gather*}
\caption{Definition of $\mathbf{Fresh} \ -$}
\label{fig:fresh}
\end{figure}

\begin{figure}[H]
  \footnotesize
  \begin{gather*}
  \begin{aligned}
  &\mathbf{Add} \ - \ \mathbf{to} \ -: (\Phi;\Sigma;\omega) \to (\Phi;\Sigma;\omega) \to (\Phi;\Sigma;\omega) \\
  &\mathbf{Add} \ (\Phi_m;\Sigma_m;\omega_m) \ \mathbf{to} \ (\Phi_n;\Sigma_n;\omega_n) = \\
  &\quad \Sigma_r := \cdot \\
  &\quad \mathbf{for} \ (x:\omega_x) \ \mathbf{in} \ \Sigma_m \\
  &\quad \quad \Sigma_r := \Sigma_r, x: \mathbf{addType}(\Sigma_m(x);\Sigma_n(x)) \\
  &\quad \mathbf{return} \ (\Phi_m,\Phi_n;\Sigma_r;\mathbf{addType}(\omega_m;\omega_n)) \\
  \\
  &\mathbf{addType}: (t;t) \to t\\
  &\mathbf{addType}(\tau^m_{\psi_m};\tau^n_{\psi_n}) = \mathbf{addType}(\tau^m;\tau^n)_{\bunch{\psi_m, \psi_n}} \\
  &\mathbf{addType}(\tau^m_{\psi_m};\tau^n) = \mathbf{addType}(\tau^m;\tau^n)_{\psi_m} \\
  &\mathbf{addType}(\tau^m;\tau^n_{\psi_n}) = \mathbf{addType}(\tau^m;\tau^n)_{\psi_n} \\
  &\mathbf{addType}(\langle \rho_m \rangle;\langle \rho_n \rangle) = \langle \mathbf{addType}(\rho_m;\rho_n) \rangle \\
  &\mathbf{addType}(\nu \ \keyword{as} \ \langle \rho_m \rangle;\nu \ \keyword{as} \ \langle \rho_n \rangle) = \nu \ \keyword{as} \ \langle \mathbf{addType}(\rho_m;\rho_n) \rangle \\
  &\mathbf{addType}(\cdot;\cdot) = \cdot \\
  &\mathbf{addType}(\nu;\nu) = \nu \\
  &\mathbf{addType}((\ell: \tau_m \mid \rho_m);(\ell: \tau_n \mid \rho_n)) = (\ell: \mathbf{addType}(\tau_m;\tau_n) \mid \mathbf{addType}(\rho_m;\rho_n)) \\
  &\mathbf{addType}(();()) = () \\
  &\mathbf{addType}((\tau_m, \tau_n);(\tau_j, \tau_k)) = (\mathbf{addType}(\tau_m;\tau_j), \mathbf{addType}(\tau_n;\tau_j)) \\
  &\mathbf{addType}(\sRes{\varphi_m}{\tau_m};\sRes{\varphi_n}{\tau_n}) = \sRes{\bunch{\varphi_m, \varphi_n}}{\mathbf{addType}(\tau_m;\tau_n)} \\
  &\mathbf{addType}(\sFun{\varphi_m}{\tau_m}{\tau_n};\sFun{\varphi_j}{\tau_j}{\tau_k}) = \sFun{\bunch{\varphi_m, \varphi_j}}{\mathbf{addType}(\tau_m;\tau_j)}{\mathbf{addType}(\tau_n;\tau_k)} \\
  &\mathbf{addType}((\sFun{\varphi_m}{\tau_m}{\tau_n}) \scFun \omega_m;(\sFun{\varphi_j}{\tau_j}{\tau_k}) \scFun \omega_j) = \mathbf{addType}(\sFun{\varphi_m}{\tau_m}{\tau_n};\sFun{\varphi_j}{\tau_j}{\tau_k}) \scFun \mathbf{addType}(\omega_m;\omega_j)
  \end{aligned}
\end{gather*}
\caption{Definition of $\mathbf{Add} \ - \ \mathbf{to} \ -$}
\label{fig:add}
\end{figure}

\newpage

\section{Proofs}
In the following proofs, we write $mems(\Phi;\omega)$ (defined in \Cref{fig:memes}) for the set of types traced by a trace member, extracted from a single-traced type $\omega$ in the tracing environment $\Phi$. The parameter $\Phi$ is omitted if it is clear from the context. It can be easily generalized to be used on environments.

\begin{figure}[H]
  \footnotesize
  \begin{gather*}
  \begin{aligned}
  &mems: (\Phi;t) \to \{\tau\} \\
  &mems(\alpha \cstOf \bunch{\beta, \phi};\tau_{\bunch{\beta}}) = \{\tau_{\bunch{\beta}}\} \\
  &mems(\Phi;\nu) = \emptyset \\
  &mems(\Phi;\langle \rho \rangle) = mems(\Phi;\rho) \\
  &mems(\Phi;()) = \emptyset \\
  &mems(\Phi;(\tau_m, \tau_n)) = mems(\Phi;\tau_m) \cup mems(\Phi;\tau_n) \\
  &mems(\Phi;\cdot) = \emptyset \\
  &mems(\Phi;(\ell : \tau \mid \rho)) = mems(\Phi;\tau) \cup mems(\Phi;\rho) \\
  &mems(\Phi;\sRes{\varphi}{\tau}) = mems(\Phi;\tau) \\
  &mems(\Phi;\sFun{\varphi}{\tau_p}{\tau_e}) = mems(\Phi;\tau_p) \cup mems(\Phi;\tau_e) \\
  &mems(\Phi;(\sFun{\varphi}{\tau_p}{\tau_e}) \scFun \omega) = mems(\Phi;\sFun{\varphi}{\tau_p}{\tau_e}) \cup mems(\Phi;\omega)
  \end{aligned}
  \end{gather*}
\caption{Definition of $mems$}
\label{fig:memes}
\end{figure}

\printProofs

\end{document}